\pdfoutput=1

\documentclass[11pt]{article}
\newcommand{\editcolor}{\color{black}}
\newcommand{\feditcolor}{\color{black}}
\usepackage[final]{acl}

\usepackage{amsmath,amsfonts,bm}

\def\eqref#1{equation~\ref{#1}}

\def\1{\bm{1}}

\DeclareMathAlphabet{\mathsfit}{\encodingdefault}{\sfdefault}{m}{sl}
\SetMathAlphabet{\mathsfit}{bold}{\encodingdefault}{\sfdefault}{bx}{n}

\DeclareMathOperator*{\argmax}{arg\,max}

\usepackage[ruled,vlined,noend,linesnumbered]{algorithm2e}
\usepackage{multirow}
\usepackage{times}
\usepackage{latexsym}
\usepackage{amssymb}
\usepackage{makecell}
\usepackage{arydshln}  
\usepackage[T1]{fontenc}

\usepackage[utf8]{inputenc}

\usepackage{microtype}
\usepackage{makecell} 
\usepackage{inconsolata}

\usepackage{graphicx}
\usepackage{tcolorbox}
\usepackage{float}
\usepackage{kotex}

\title{{\editcolor HELIOS: Harmonizing Early Fusion, Late Fusion, and LLM Reasoning for Multi-Granular Table-Text Retrieval}}

\author{
Sungho Park \\
POSTECH, Republic of Korea \\
\texttt{shpark@dblab.postech.ac.kr}
\And
Joohyung Yun \\
POSTECH, Republic of Korea \\
\texttt{jhyun@dblab.postech.ac.kr}
\AND
Jongwuk Lee \\
Sungkyunkwan University, Republic of Korea \\
\texttt{jongwuklee@skku.edu}
\And
Wook-Shin Han\thanks{\ \ Corresponding author.} \\
POSTECH, Republic of Korea \\
\texttt{wshan@dblab.postech.ac.kr}
}

\begin{document}
\maketitle

\begin{abstract}
\vspace{-1.5mm}

Table-text retrieval aims to retrieve relevant tables and text to support open-domain question answering. 
Existing studies use either early or late fusion, but face limitations. 
Early fusion pre-aligns a table row with its associated passages, forming ``stars," which often include irrelevant contexts and miss query-dependent relationships. 
Late fusion retrieves individual nodes, dynamically aligning them, but it risks missing relevant contexts. 
Both approaches also struggle with advanced reasoning tasks, such as column-wise aggregation and multi-hop reasoning. 
To address these issues, we propose \texttt{HELIOS}, which combines the strengths of both approaches. 
First, the \emph{edge-based bipartite subgraph retrieval} identifies finer-grained edges between table segments and passages, effectively avoiding the inclusion of irrelevant contexts. 
Then, the \emph{query-relevant node expansion} identifies the most promising nodes, dynamically retrieving relevant edges to grow the bipartite subgraph, minimizing the risk of missing important contexts. 
Lastly, the \emph{star-based LLM refinement} performs logical inference at the star graph level rather than the bipartite subgraph, supporting advanced reasoning tasks. 
Experimental results show that \texttt{HELIOS} outperforms state-of-the-art models with a significant improvement up to 42.6\% and 39.9\% in recall and nDCG, respectively, on the \texttt{OTT-QA} benchmark.
\end{abstract}
\section{Introduction}
\vspace{-1.5mm}

\begin{figure}[t]
    \includegraphics[width=\columnwidth]{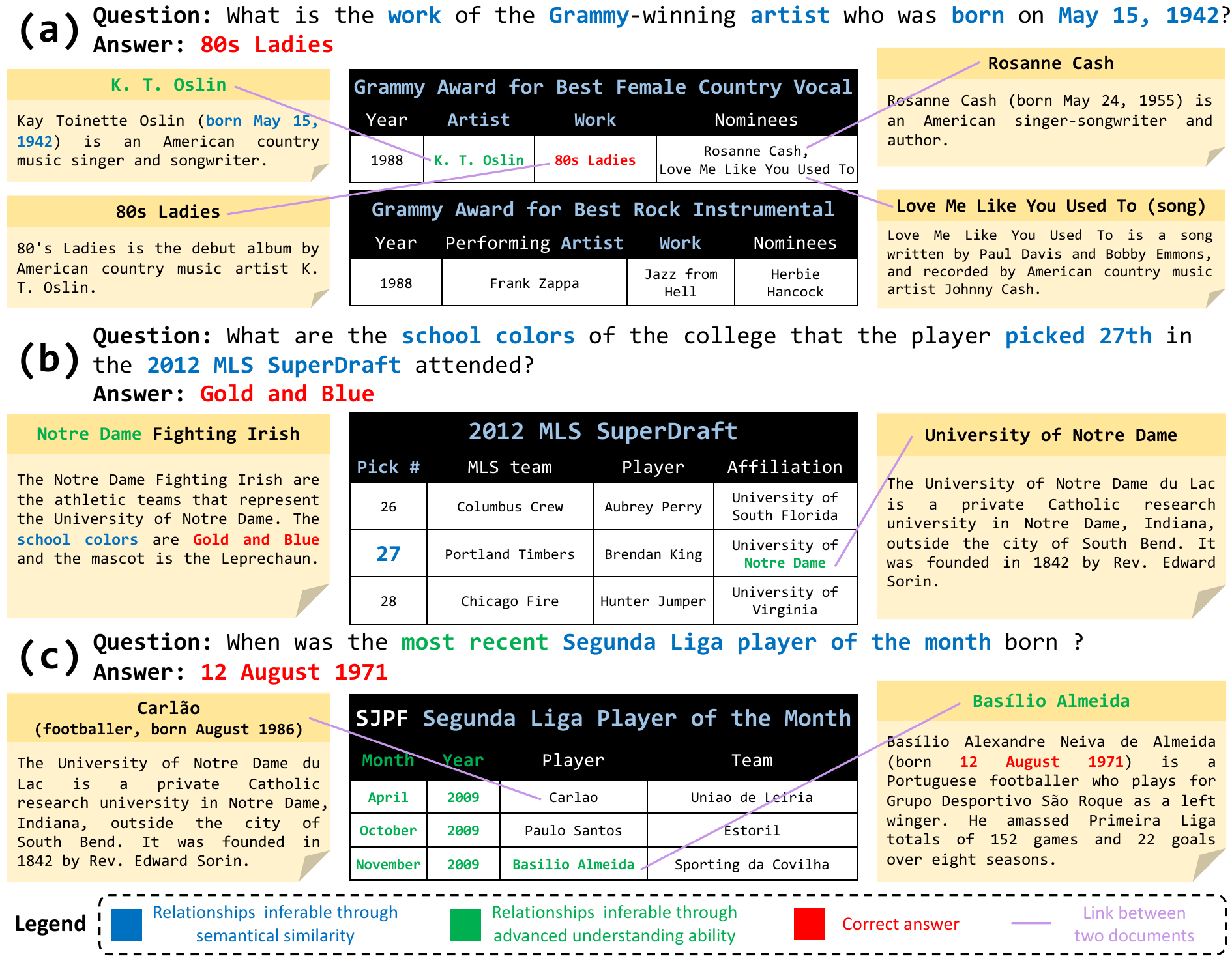}
    \vspace{-3mm}
    \caption{
        Simplified examples of three cases where existing methods struggle to retrieve the question-related documents correctly.
        (a) Inadequate granularity of retrieval units leading to inaccurate retrieval results.
        (b) Entity linking results cannot estimate essential query-aware relationships.
        (c) Inability of advanced reasoning such as table aggregation and multi-hop reasoning.
    }
    \vspace{-5mm}
\label{fig:motivation}
\end{figure}

Open-domain question answering (ODQA) over tables and text is important as it leverages the complementary strengths of structured and unstructured data.
Tables offer vast amounts of related facts but lack diversity, while text provides broader contextual information~\citep{Hybridqa, OTT-QA}, making the integration of both modalities essential.
Table-text retrieval plays a crucial role in ODQA by retrieving relevant tables and text to support retriever-reader systems ~\citep{OTT-QA, COS, CORE}.

Despite its importance, table-text retrieval is challenging due to the need to bridge structured tables and unstructured passages.
Tables encode information in rows and columns, requiring structural understanding, while passages follow a narrative format. Effective retrieval demands resolving multi-hop relationships across these distinct formats.

Existing methods have achieved some success by employing either \textit{early} or \textit{late fusion} techniques in their top-$k$ retrieval.
The \emph{early fusion} attempts to reduce the search space by grouping relevant documents before a query is presented. 
It pre-aligns a table row with associated passages via entity linking, creating a \emph{fused block} as the retrieval unit~\citep{OTT-QA, OTTeR, DoTTeR}. 
In contrast, the late fusion aligns relevant table rows and passages dynamically using query-based similarity matching after the query is given.
It returns a ranked sequence of evidence chains, where an \emph{evidence chain} refers to a pair consisting of a table row and a passage ~\citep{CORE, COS}.

However, the existing studies have several significant limitations.

\textbf{(1) Inadequate granularity of retrieval unit.} 
Early fusion strategy~\citep{OTT-QA, OTTeR, DoTTeR} constructs retrieval units independently of the query, often including query-irrelevant passages, which distorts similarity calculations between fused blocks and queries.
For example in Figure~\ref{fig:motivation}(a), entity linking connects the \texttt{Grammy} \texttt{Award} \texttt{for} \texttt{Best} \texttt{Female} \texttt{Country} \texttt{Vocal} table with four surrounding passages, even though only the information on the \texttt{K. T. Oslin} is relevant (Figure~\ref{fig:motivation}(a)).
In the late fusion strategy~\citep{CORE, COS}, retrieving a single table segment or passage may be partially relevant to a query, incurring the risk of retrieving incorrect tables.
For instance, during the first iteration of retrieval, the system might retrieve the \texttt{Grammy} \texttt{Award} \texttt{for} \texttt{Best} \texttt{Rock} \texttt{Instrumental} table instead of the correct one. 
Both tables share overlapping terms such as \texttt{Grammy}, \texttt{Artist}, and \texttt{Work}, causing ambiguity in target identification.

\textbf{(2) Missing query-dependent relationships.} The early fusion strategy~\citep{OTT-QA, OTTeR, DoTTeR} relies on entity linking to predefine relationships between tables and passages, failing to capture query-dependent links that might contain the information necessary to answer the query.
For instance, in Figure~\ref{fig:motivation}(b), the table \texttt{2012} \texttt{MLS} \texttt{SuperDraft} is early fused with the entity \texttt{University} \texttt{of} \texttt{Notre} \texttt{Dame}. 
However, when the question specifies the information about \texttt{school} \texttt{colors}, it should be linked to the \texttt{Notre} \texttt{Dame} \texttt{Fighting} \texttt{Irish} passage.

\textbf{(3) Lack of advanced reasoning.} 
Queries that require complex reasoning, such as multi-hop or column-wise aggregation, often demand advanced logical inference beyond simple semantic similarity with the query. 
Since previous approaches~\citep{OTT-QA, OTTeR, DoTTeR, COS, CORE} rely on semantic similarity, they might fail to retrieve rows or passages identifiable through logical inference.
For example, in Figure~\ref{fig:motivation}(c), the query involves understanding the \texttt{most} \texttt{recent} \texttt{Segunda} \texttt{Liga} \texttt{Player} \texttt{of} \texttt{the} \texttt{Month} is \texttt{Basilio} \texttt{Almeida}, where the row with the latest \texttt{Year} and \texttt{Month} combination has to be inferred. 

To systematically address these limitations, we first formalize the terms proposed in previous studies using a \emph{bipartite graph}, where table segments and passages are represented as two disjoint sets of nodes, and the links between them are represented as edges. 
Therefore, the term \emph{fused block} used in the early fusion strategy \citep{OTT-QA, OTTeR, DoTTeR} can be represented as a star \citep{GraphTheory} centered on a node of type \texttt{table segment}, with connected nodes of type \texttt{passage}. 
Similarly, the \emph{evidence chain} used in the late fusion strategy \citep{CORE, COS} corresponds to an edge connecting a pair of nodes: one of type \texttt{table segment} and one of type \texttt{passage}.

Based on this formalization, we propose \texttt{HELIOS}, a novel graph-based retrieval consisting of three stages: early fusion, late fusion, and LLM reasoning. 
Specifically, \texttt{HELIOS} adopts the following three key ideas:

\textbf{(1) Combined usage of early and late fusion.}
We selectively leverage the advantages of both early fusion and late fusion. 
Early fusion pre-aligns tables with related passages to mitigate the risk of retrieving incomplete or partially relevant information inherent in late fusion, while late fusion dynamically resolves document relationships to address early fusion's reliance on predefined links.

\vspace{-1mm}
\textbf{(2) Graph refinement.} 
We leverage LLMs to perform further advanced reasoning over the retrieved graph, enabling deeper logical inference beyond simple semantic similarity.
For instance, in Figure~\ref{fig:motivation}(c), when the \texttt{SJPF} \texttt{Segunda} \texttt{Liga} \texttt{Player} \texttt{of} \texttt{the} \texttt{Month} table is retrieved, the LLM can perform aggregation to identify the most recent player and conduct multi-hop reasoning to select the corresponding passage for \texttt{Basilio} \texttt{Almeida}.

\textbf{(3) Granularity determination for each retrieval stage.}
In our retrieval pipeline, each stage early fusion, late fusion, and graph refinement serves a distinct purpose, necessitating the precise determination of the appropriate operational units for each.
For the early fusion stage, we propose an {\editcolor edge-level, multi-vector-based retrieval}, striking a balance between eliminating irrelevant contexts in star graph retrieval and avoiding the partial information problem in node-based retrieval.
In the late fusion stage, we set the unit as an individual node.
We identify query-relevant nodes within the graph produced by the early fusion stage so that we can design the late fusion process to expand the graph using only nodes closely aligned with the query context.
This approach mitigates the challenge where the earlier stage may retrieve nodes irrelevant to the query.
Finally, the graph refinement stage presents an expanded graph from late fusion to the LLM, reducing hallucination risks by decomposing the graph into smaller star graphs.

{\feditcolor
We evaluate \texttt{HELIOS} and its competitors on the \texttt{OTT-QA} \citep{OTT-QA} and \texttt{MultimodalQA (MMQA)} \citep{MultiModalQA} datasets.
}
Experimental results demonstrate that \texttt{HELIOS} significantly outperforms SOTA systems.
\vspace{-1mm}
\section{Related Work}
\vspace{-1mm}

\vspace{-1mm}
\subsection{Open-domain Question Answering}
\vspace{-1mm}

Open-Domain Question Answering (ODQA) aims to answer factual questions using a large knowledge corpus (\citealp{zhang-etal-2023-survey-efficient}). Standard benchmarks like Natural Questions \citep{NaturalQuestions}, TriviaQA \citep{TriviaQA}, and SearchQA \citep{SearchQA} focus on single-hop queries, where answers reside within a single passage of unstructured text. 
Further advances were shown by HotpotQA (\citealp{HotpotQA}) and WikiHop (\citealp{WikiHop}), presenting challenging queries that require multi-hop reasoning across multiple passages.
However, these benchmarks consider only unstructured text and do not address multi-hop reasoning over both structured tables and unstructured passages, which is essential in table-text retrieval tasks.
OTT-QA \citep{OTT-QA} is the first ODQA benchmark designed to support multi-hop reasoning between tables and text, requiring retrieval and reasoning across both modalities.

\vspace{-1mm}
\subsection{Table-Text Retrieval}
\vspace{-1mm}

Table-text retrieval methods can be categorized into early fusion and late fusion approaches. These terms, originally used in multi-modal tasks such as image-sentence retrieval, describe whether different modalities are encoded jointly or separately \citep{wang2022comprehensive, snoek2005early, gadzicki2020early}.
Similarly, in table-text retrieval, early fusion and late fusion approaches differ in whether tables and text are linked before or after the retrieval process~\citep{DoTTeR}.

\emph{Early fusion} approaches \citep{OTT-QA, OTTeR, DoTTeR} pre-link table rows with associated passages via entity linking, forming fused blocks as retrieval units.
While convenient, this approach has two drawbacks:
(i) Fused blocks often include query-irrelevant passages, causing noisy retrieval and information loss in block-level embedding. 
(ii) Offline pre-linked blocks cannot adapt to query-dependent relationships discovered during retrieval (Figure~\ref{fig:motivation}(b)).
To address (i), we adopt edges—a finer-grained retrieval unit—and employ late interaction during retrieval to minimize information loss and avoid large, noisy blocks. 
For (ii), we propose query-relevant node expansion to incorporate relationships that emerge in a specific query's context.

\emph{Late fusion} approaches \citep{CORE, COS} dynamically form table-passage connections online. 
Although more flexible, they must consider many table-passage combinations, typically handled by beam search, which can cause error propagation.
Our approach mitigates this using edge-based late interaction retrieval, which captures richer contextual relationships by pre-linking table segments with passages offline, enabling more accurate seed document retrieval.


\emph{Both} early and late fusion approaches rely primarily on semantic similarity for retrieval, limiting their ability to retrieve table segments and passages requiring advanced reasoning (e.g., column-wise aggregation, multi-hop inference), as shown in Figure~\ref{fig:motivation}(c).
To address this, we propose a star-based LLM refinement step, leveraging LLMs for logical inference to refine retrieval results.

{\feditcolor
DRAMA \citep{DRAMA} and HOLMES \citep{HOLMES} also adopt graph-based multi-hop QA, similar to our approach. 
However, both methods operate in a constrained setting with given evidence, unlike our open-domain setup.
GTR \citep{GTR} and MGNETS \citep{MGNETS} improve table encoding using graph-based methods, addressing a problem orthogonal to our focus. 
In contrast, our work centers on retrieving both tables and text based on their semantic relationships.
}
\vspace{-1mm}
\section{Preliminaries}
\vspace{-1mm}

\subsection{Problem Formulation}
\label{sec:problem_formulation}
\vspace{-1mm}

Table-text retrieval is involved from a retrieval \textit{corpus} $\mathcal{C}$, which comprises two distinct sets: a collection of passages $\mathcal{C}_P = \{P^{(1)}, \dots, P^{(n)}\}$ and a collection of tables $\mathcal{C}_T = \{T^{(1)}, \dots, T^{(m)}\}$. 
A \textit{passage} is defined as a sequence of tokens $P$, representing unstructured text. 
A \textit{table} is a structured matrix $T$, consisting of cells $T_{i, j}$, where $i$ and $j$ indicate the row index and the column index, respectively. 
Each cell $T_{i, j}$ may contain a number, date, phrase, or sentence.
We define a document as either a passage or a table.
Given a query $q$, the objective of table-text retrieval is to retrieve from corpus $\mathcal{C}$ a ranked list of documents such that the document containing the answer span $a$ is positioned among the top results.

We split a table into multiple table segments, as commonly used in existing studies. 
Since a single table can easily exceed the token limits of language models, a table $T$ is combined with its header to form a list of table segments $T = [S^{(1)}, \dots, S^{(m')}]$ (\citealt{OTT-QA}). 
This process results in (i) a corpus $\mathcal{C}$ composed of table segments $\mathcal{C}_S$ and passages $\mathcal{C}_P$ (i.e., $\mathcal{C} = \mathcal{C}_S \cup \mathcal{C}_P$) and (ii) a mapping $\mathcal{M}: \mathcal{C}_S \rightarrow \mathcal{C}_T$ to associate table segments with their original table.

\vspace{-1mm}
\subsection{Table-Text Retrieval as Graph Retrieval}
\label{sec:table_text_retrieval_as_bipartite_graph_retrieval}
\vspace{-1mm}

We adopt a graph representation, denoted by $G = (V, E, \Phi, \Gamma, \Lambda)$, to generalize various methods used in existing studies. 
Here, $V$ is the set of vertices corresponding to a table segment or a passage, and $E$ is the set of edges representing relationships between (table segment, passage) pairs. 
The mapping $\Phi: V \rightarrow \{\texttt{table}\ \texttt{segment}, \texttt{passage}\}$ maps each node to its type, while $\Gamma$ maps a node to its attributes, such as the text of a passage or the matrix of table structures. 
The mapping $\Lambda: E \rightarrow \mathbb{R}$ maps each edge to its score.

The corpus can be expressed as the initial graph $G_{init} = (V_{init}, \emptyset, \Phi, \Gamma, \Lambda_{init})$, where each node in $V_{init}$ one-to-one corresponds to a table segment or passage in $\mathcal{C}$.  
Early fusion generates table-text relationships via entity linking and updates $G_{init}$ before a query $q$ is presented.  
Given $q$, late fusion dynamically generates query-dependent table-text relationships to update $G_{init}$.  
Finally, we retrieve a query-relevant edge-scored bipartite graph $G_q = (V_q, E_q, \Phi, \Gamma, \Lambda_q)$ from $G_{init}$.  
This problem is often interpreted as ranking edges $\mathcal{E}_q$ from all possible edges, as retrieved results are fed to a reader with limited context size~\citep{CORE, COS}.  
$\mathcal{E}_q$ is typically obtained by sorting each edge $e$ in $G$ using its edge scores $\Lambda_q(e)$.  

\vspace{-1mm}
\section{Proposed Method}
\label{sec:methodology}

\vspace{-1mm}
\begin{figure*}[h!]
\centering
    \includegraphics[width=\textwidth]{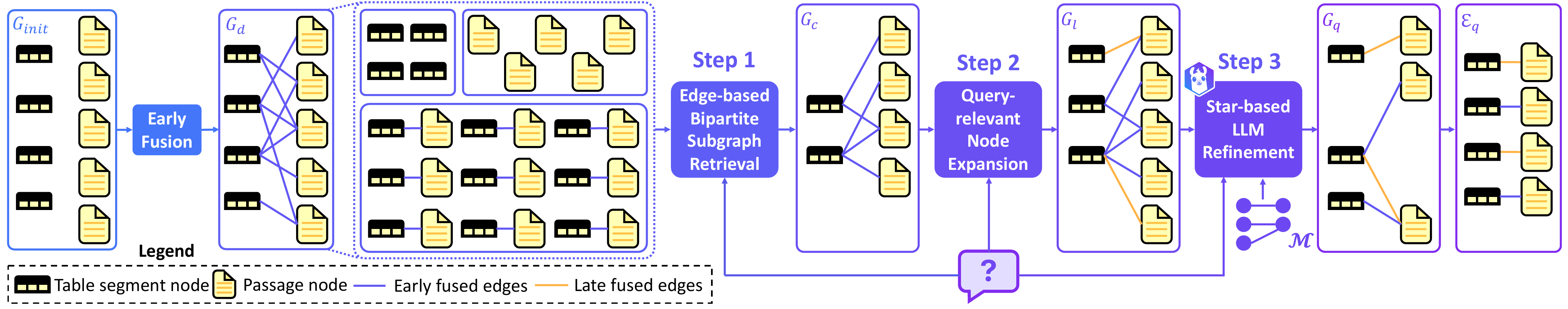}
    \vspace{-2mm}
    \caption{
        Overview of \texttt{HELIOS}:
        The initial graph $G_{init}$ is early-fused to generate a graph $G_d$.
        Each node and edge of $G_d$ are embedded.
        (1) The edges of $G_d$ are retrieved using the query $q$, then integrated into a candidate bipartite subgraph $G_c$.
        (2) The most query-relevant nodes in $G_c$ are identified as seed nodes.
        New nodes from $G_{init}$ are expanded from the seed nodes, forming an expanded graph $G_l$.
        (3) LLM performs aggregation over restored tables to identify new relevant table rows, and then eliminates irrelevant passages.
    }
    \vspace{-5mm}
\label{fig:system_overview}
\end{figure*}
We propose \texttt{HELIOS}, a novel graph-based retrieval framework that combines the strengths of early fusion, late fusion, and LLM reasoning. 
As shown in Figure~\ref{fig:system_overview}, it operates in three stages: 
(i) \textbf{Edge-based Bipartite Subgraph Retrieval} retrieves edges from a bipartite data graph constructed via early fusion and integrates them into a single bipartite subgraph. 
(ii) \textbf{Query-relevant Node Expansion} enhances the retrieved subgraph by incorporating additional nodes through further retrieval. 
(iii) \textbf{Star-based LLM Refinement} refines the expanded graph through aggregation and multi-hop reasoning using the LLM.

\vspace{-1mm}
\subsection{Edge-based Bipartite Subgraph Retrieval}
\label{sec:candidate_bipartite_graph_generation}
\vspace{-1mm}

\texttt{HELIOS} initiates its process with the retrieval of a bipartite subgraph through two key steps: \textit{early fusion} and \textit{edge retrieval}.

\textbf{(1) Early fusion}: 
This step is performed offline, before a query is given.
A bipartite data graph $G_d$ is constructed by generating edges from $G_{init}$, which initially has no edges. 
Edge generation is composed of entity recognition and entity linking, following prior methods~\citep{CORE, COS}.
Edges are generated between passage nodes and table segment nodes, resulting the generated data graph $G_d$ to be a bipartite graph.
$G_d$ is then indexed in an edge-wise manner.
Each edge $e = (S, P)$ is first linearized into a token sequence $x$.
\vspace{-1mm}
\begin{equation}    
    x = [\ Linearize(\Gamma(S));\ \Gamma(P)\ ]
    \vspace{-1mm}
\end{equation}
Then, $x$ is embedded into a sequence of vectors.
\vspace{-2mm}
\begin{equation}    
    \textbf{X} = f_{e}(x) \in \mathbb{R}^{l_x \times d}
    \vspace{-2mm}
\end{equation}
where $l_x$ represents the length of a linearized edge $x$.
{\editcolor We adopt the multi-vector encoder \texttt{ColBERTv2} (\citealp{ColBERTv2}) for $f_e$ to create fine-grained embeddings.
That is, HELIOS embeds edges, a larger unit compared to the node-level embeddings used in previous methods (\citealp{CORE, COS}). 
Since edges contain more tokens than nodes, fine-grained embeddings are essential to reduce information loss.}

\textbf{(2) Edge Retrieval}: 
When the query is given, we first identify the query-relevant edges by leveraging the semantic similarity between the query and edge embeddings.
\vspace{-2mm}
\begin{equation}
    \textbf{Q} = f_{e}(q) \in R^{l_q \times d}
    \vspace{-2mm}
\end{equation}
The similarity is then calculated between the query and each indexed edge.
\vspace{-2mm}
\begin{equation}
     sim(q, e;f_{e}) = \sum_{i=1}^{l_q} \max_{j\in [1, l_x]} \textbf{Q}_i\textbf{X}_j^T
     \vspace{-2mm}
\label{eqn:similarity_query_edge}
\end{equation}
We then select the top-$k_1$ edges that show the highest score $sim(q, e;f_{e})$, measuring query-edge alignment.
The selected edges are further passed to a fine-tuned all-to-all interaction reranker $g_{e}$, which performs a more detailed similarity evaluation.
This identifies the most contextually relevant edges, allowing us to identify the top-$k_2$ query-relevant edges $(k_2 < k_1)$.
The fine-tuning process for $f_e$ and $g_e$, including training dataset construction, is detailed in Appendix~\textsection~\ref{sec:edge_retriever_and_reranker}.

The resultant edges are integrated into the bipartite subgraph $G_c = (V_c, E_c, \Phi, \Gamma, \Lambda_c)$. 
Specifically, if there are duplicate nodes contained in the retrieved edges, those nodes are merged to form a single graph.
$G_c$ serves as the candidate bipartite subgraph, which becomes the foundation for further expansion and refinement.
We save the similarity score for each edge in the score mapping function $\Lambda_c$, which is later used to sort the edges based on their relevance to the query in Section~\ref{sec:graph_refinement_via_complex_reasoning}.


\vspace{-1mm}
\subsection{Query-relevant Node Expansion}
\label{sec:candidate_bipartite_graph_augmentation}
\vspace{-1mm}

\begin{figure*}[h!]
\centering
    \includegraphics[width=0.85\textwidth]{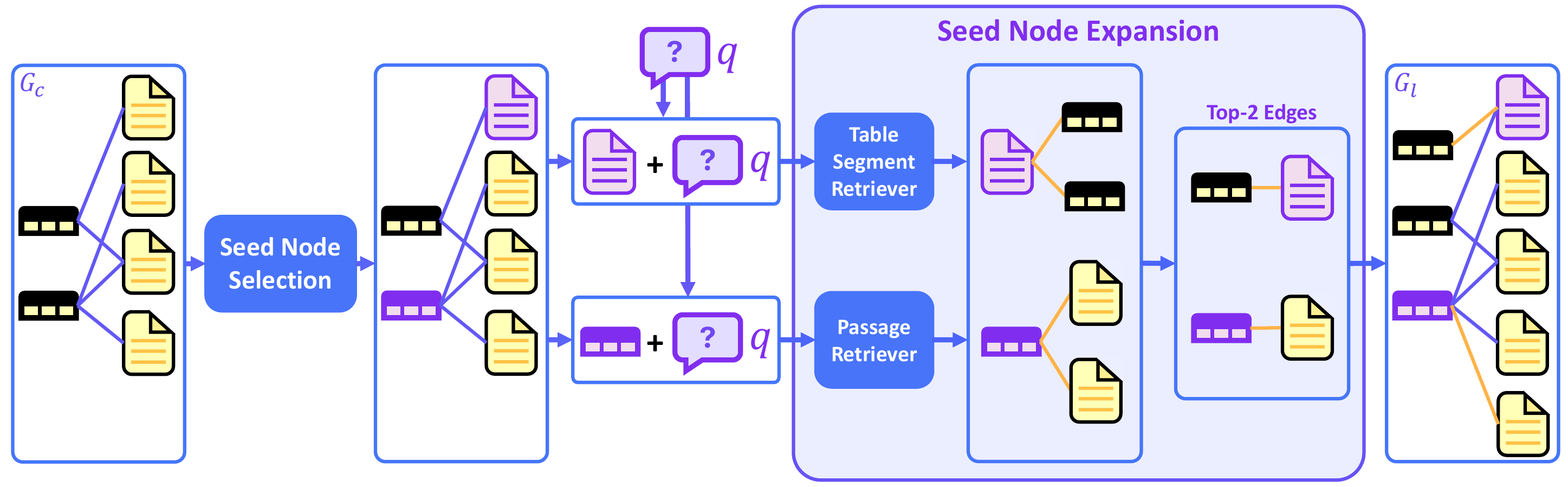}
    \caption{
        The overall procedure of query-relevant node expansion.
        The beam width $b$ is set as 2 in this example. 
        The purple-colored nodes indicate the selected seed nodes.
    }
    \vspace{-5mm}
\label{fig:query_driven_late_fusion}
\end{figure*}

Query-relevant node expansion process aims to identify additional query-relevant edges, including the edges that have not been present within $G_d$. 
We perform the expansion process at the node level, which is the most fine-grained level. 
This is to address the issue that early fusion inevitably includes query-irrelevant nodes in the candidate subgraph.
We formalize the process as finding a set of edges that meet the following objective function:
\vspace{-3mm}
\begin{equation}
    \argmax_{(u,v)\in E^*\wedge u \in V_c} p(u,v|q) =  p(v|u,q)p(u|q)
    \vspace{-2mm}
\label{eqn:query_driven_late_fusion_objective}
\end{equation}
Here, $u$ represents a node in the candidate graph $G_c$, and $v$ represents a node adjacent to $u$ in the complete bipartite graph $G^*$. 
The complete bipartite graph $G^* = (\mathcal{C}_{init}, E^*, \Phi, \Gamma, \Lambda_{init})$ contains all possible edges between table segments and passages. 
Naively solving this objective requires calculating the similarity score between the query and every possible edge in the complete graph, which is infeasible. 
To address this, we adopt a beam search approach and model it as a two-step process, as illustrated in Figure~\ref{fig:query_driven_late_fusion}.

\textbf{(1) Seed node selection}: 
This step composes the first iteration of the beam search, corresponding to finding the set of nodes that show the highest $p(u|q)$.
We calculate $p(u|q)$ for each $u \in V_c$ to identify the top-$b$ (i.e., beam width) nodes that contain the most relevant information to the query.
The probability $p(u|q)$ is determined by calculating the semantic similarity between the query and each node $u$ in $G_c$, then normalizing the scores using a softmax function. 
This similarity is computed through a fine-tuned all-to-all interaction-based node reranker $g_{n}$.

\textbf{(2) Seed node expansion}: 
This step composes the second iteration of the beam search, aiming to find the set of edges $(u, v)$ which show the highest $p(u, v|q)$.
It is done by computing $p(v|u,q)$ for each node $v$ connected to a seed node $u$ in the complete bipartite graph $G^*$.
This conditional probability is calculated using the expanded query retrieval technique (\citealp{expandedqueryretrieval}).
In this technique, the score function is expressed based on the expanded query as $sim([q;\Gamma(u)], v)$, and it is calculated by two late interaction models: 
$sim([q;\Gamma(u)], v; f_{P \rightarrow S})$
 for a table-segment-typed expanding node and a passage-typed seed node, and $sim([q;\Gamma(u)], v; f_{S \rightarrow P})$ for the opposite.
The calculated scores are normalized using a softmax function to compute $p(v|u,q)$.
Finally, $p(u, v|q)$ is calculated for the $(u, v)$ pairs following Equation~\ref{eqn:query_driven_late_fusion_objective}, and the top-$b$ edges with the highest scores are finally selected.
The $(u, v)$ pairs are added to $E_c$ along with the nodes $v$ to $V_c$, forming the expanded bipartite graph $G_l = (V_l, E_l, \Phi, \Gamma, \Lambda_l)$. 
The scores for each new edge are calculated using the reranker $g_e$ as used in \textsection~\ref{sec:candidate_bipartite_graph_generation}.
Detailed explanation for fine-tuning $g_n$, $f_{P\rightarrow S}$ and $f_{S\rightarrow P}$ can be found in Appendix~\textsection~\ref{sec:node_reranker} and \textsection~\ref{sec:expanded_query_retrievers}.


\vspace{-2mm}
\subsection{Star-based LLM Refinement}
\label{sec:graph_refinement_via_complex_reasoning}
\vspace{-1.5mm}


The main goal of this step is to retrieve query-relevant information which is challenging to find using semantic similarity alone, by leveraging the logical inference capabilities of LLMs.
Selecting the optimal format or unit for presenting the graph $G_l$ to the LLM is non-trivial.
We explored two approaches: (i) providing the entire graph $G_l$ in a single prompt to return the relevant set of nodes and (ii) decomposing $G_l$ into star graphs, with each star graph generating its own set of relevant nodes.
The latter approach proved to be 12.4\% more effective, leading us to adopt star graphs as the unit for logical inference (\textsection~\ref{sec:exp_granularity_analysis}). 
An overview of the process and few-shot prompts is in Appendix~\ref{sec:star_based_llm_refinement_supp} and \textsection~\ref{sec:Prompts}.
The process consists of two phases: \emph{column-wise aggregation} and \emph{passage verification}.



\textbf{(1) Column-wise Aggregation}: 
This step aims to infer the correct result rows for table aggregation operations, as exemplified in Figure~\ref{fig:motivation}(c), making it possible that the corresponding rows exist in $G_l$. 
Since not every query requires aggregation, we first prompt the LLM to determine whether the input query necessitates an aggregation operation. 
If the query is classified as an aggregation query, it first restores the original from each table segment using mapping function $\mathcal{M}$. 
The restored tables are then provided to the LLM in the format of star graph along with the query. 
The LLM performs the aggregation and returns the rows corresponding to the aggregation result.
The returned rows (i.e., table segments) are subsequently added to $G_l$ along with their associated passages to generate $G_a$.

\textbf{(2) Passage Verification}: 
The edges of $G_a$ may contain lots of hard negatives, as they comprise the edges retrieved from the edge retrieval, node expansion and column-wise aggregation step.
The passage verification step aims to identify the passages relevant to answering the query.
Similar to the column-wise aggregation step, we provide $G_a$ to the LLM in the form of star graphs, units that contain multi-hop relationships.
The LLM performs a binary verification to determine whether each edge is relevant to the query, without recalculating their scores. 
As a result, query-irrelevant edges are removed, yielding a refined edge-scored graph $G_q$. 

We transform the graph $G_q$ into a sequence of edges, as our reader requires a serialized token sequence as input.
We use the mapping function $\Lambda_c$ to sort each edge $e$ by its similarity to the query $\Lambda_c(e)$, and the top $k$ edges are returned.

\vspace{-1.5mm}
\section{Experiments}
\label{sec:experiments}
\vspace{-1.5mm}



\textbf{Hardware and Software Settings.}
We conducted our experiments on a machine with AMD EPYC 7313 CPU and 2TB of RAM with the OS of Rocky Linux release 8.7 and 4 A100-80GB GPUs.

\textbf{Competitors.}
\texttt{HELIOS} is compared with the SOTA methods. 
The early fusion methods include \texttt{Fusion-Retriever} (\citealp{OTT-QA}), \texttt{OTTeR} (\citealp{OTTeR}), and \texttt{DoTTeR} (\citealp{DoTTeR}). 
The late fusion approaches include \texttt{Iterative-Retriever} (\citealp{OTT-QA}), \texttt{CORE} (\citealp{CORE}), and \texttt{COS} (\citealp{COS}).
{
\feditcolor
Additionally, we include \texttt{HOLMES}~(\citealp{HOLMES}), a graph-based multi-hop QA method originally designed for a distractor setting, in which a question-answering system is presented with exactly 10 candidates before producing an answer.
}

\textbf{Datasets.}
We conduct experiments on two datasets: \texttt{OTT-QA} (\citealp{OTT-QA}) and \texttt{MultimodalQA (MMQA)} (\citealp{MultiModalQA}). 
\texttt{OTT-QA} serves as the primary benchmark, as it is the only dataset designed for open-domain QA over tables and text, comprising 400K tables, 5M passages, and 42K training QA pairs, with 2K QA pairs each for development and testing.
\texttt{MMQA} is a multi-hop QA dataset spanning images, passages, and tables. 
While not fully aligned with our task, we use it as a supplementary benchmark to assess generalizability. 
We exclude image-based questions and conduct experiments in an open setting using its full corpus (10K tables, 210K passages, and 1.3K QA pairs) without reference candidates. More detailed experimental settings are in Appendix~\textsection~\ref{sec:experiment_supp}.

\vspace{-1mm}
\subsection{Main Results}
\label{sec:exp_main_results}
\vspace{-1mm}

\begin{table*}[ht]
    \centering
    \resizebox{0.9\textwidth}{!}{%
    \begin{tabular}{l|ccccc|c|c}
        \hline
        \textbf{Model}  & \textbf{AR@2}  & \textbf{AR@5} & \textbf{AR@10} & \textbf{AR@20} & \textbf{AR@50} & \textbf{nDCG@50} & \textbf{HITS@4k} \\
        \hline
        \texttt{Iterative Retriever}     & --            & --            & --            & --            & --            & --            & 27.2          \\
        \texttt{Fusion Retriever}        & --            & --            & --            & --            & --            & --            & 52.4          \\
        \texttt{OTTeR}$^{\dagger}$       & 31.4          & 49.7          & 62.0          & 71.8          & 82.0          & 25.9          & 70.1          \\
        \texttt{DoTTeR}$^{\dagger}$      & 31.5          & 51.0          & 61.5          & 71.9          & 80.8          & 26.7          & 70.3          \\
        \texttt{CORE}$^{\dagger}$        & 35.3          & 50.7          & 63.1          & 74.5          & 83.1          & 25.4          & 77.2          \\
        \texttt{COS}$^{\dagger}$         & 44.4          & 61.6          & 70.8          & 79.5          & 87.8          & 33.6          & 81.8          \\
        \texttt{COS w/ ColBERT \& bge}$^{\dagger}$         & 49.6          & 68.2          & 78.7          & 85.0          & 91.7          & 36.5          & 85.9          \\
        \texttt{DoTTeR + COS + LLM}$^{\dagger}$         & 50.0          & 62.4          & 70.0          & 76.2          & 84.7          & 34.7          & --          \\
        \hline
        \texttt{HELIOS}           & \textbf{63.3} & \textbf{76.7} & \textbf{85.0} & \textbf{90.4} & \textbf{94.2} & \textbf{47.0} & \textbf{91.8} \\
        \hline
    \end{tabular}
    }
    \vspace{-3mm}
    \caption{
        Retrieval accuracy on  \texttt{OTT-QA}'s dev set. Results marked with ${\dagger}$ indicate reproduced values.
    }
    \label{tab:ottqa_retrieval_accuracy} 
\end{table*}

\begin{table}[t]
    \centering
    \setlength{\tabcolsep}{2pt}
    \resizebox{\columnwidth}{!}{%
        \begin{tabular}{l|ccccc|cc}
        \hline
        \textbf{Model}            & \textbf{AR@2}   & \textbf{AR@5}   & \textbf{AR@10}  & \textbf{AR@20}  & \textbf{AR@50} & \textbf{EM} & \textbf{F1}  \\
        \hline
        \texttt{COS}$^{\dagger}$ & 50.7            & 59.7            & 67.1            & 72.4            & 79.5            & 	54.4 & 63.7\\
        \texttt{HELIOS}   & \textbf{70.5}   & \textbf{77.8}   & \textbf{81.0}   & \textbf{82.6}   & \textbf{86.2} & \textbf{59.6} & \textbf{69.1} \\
        \hline
        \end{tabular}
    }
    \vspace{-3mm}
    \caption{
        Retrieval and end-to-end QA accuracy on the \texttt{MMQA} dev set.
    }
    \label{tab:mmqa_retrieval_accuracy} 
\end{table}

We evaluate retrieval accuracy using top-$k$ Answer Recall (AR@$k$), nDCG@$k$, and Hits@4K, alongside end-to-end performance metrics: Exact Match (EM) and F1 score. 
AR@$k$ measures the proportion of queries where the correct answer appears in the top-$k$ retrieved edges \citep{COS}, while nDCG@$k$ measures the ranking quality considering both the relevance and the position of the retrieved edges. 
Hits@4K evaluates whether the answer span remains within the top 4096 tokens after linearizing ranked edges \citep{OTT-QA}.
To analyze the impact of retrieval accuracy on question-answering performance, we conduct end-to-end evaluations using EM and F1 scores.
We select $k \in \{2, 5, 10, 20, 50\}$ based on evaluation protocols of state-of-the-art early and late fusion models \citep{DoTTeR, COS}.
If $\mathcal{E}_q$ contains fewer edges than the retrieval target, we incorporate edges removed during the star-based LLM refinement stage to ensure a comprehensive assessment.

Table~\ref{tab:ottqa_retrieval_accuracy} shows the retrieval accuracy of \texttt{HELIOS} on \texttt{OTT-QA}'s development set.
\texttt{HELIOS} consistently outperforms all competitors on AR@$k$ across different $k$ values.
It outperforms the state-of-the-art \texttt{COS} model by an average of 19.0\% in AR, with the performance gap widening as $k$ decreases. 
At $k=2$, \texttt{HELIOS} achieves as much as 42.6\% higher answer recall than \texttt{COS}. 
This improvement is further reflected in nDCG@50, where \texttt{HELIOS} exhibits a 39.9\% gain.
Additionally, the Hits@4K metric shows a 12.2\% improvement over \texttt{COS}.
We report an enhanced version of \texttt{COS}, denoted as \texttt{COS w/ ColBERT \& bge}, which incorporates \texttt{ColBERT} retrievers and a \texttt{bge} reranker. 
Since \texttt{HELIOS} employs late-interaction retrieval, which generally outperforms single-embedding retrievers, we ensure \texttt{COS} uses the same retriever and reranker for a fair comparison. 
While this modification improves nDCG@50 by 8.6\% over the original \texttt{COS}, \texttt{HELIOS} still outperforms the enhanced version of \texttt{COS} by a substantial margin of 28.8\%.
{\feditcolor
We also evaluate a baseline that stacks \texttt{DoTTeR} (the SOTA in early fusion), COS (the SOTA in late fusion), and a full-graph prompting approach for LLM-based reasoning to test whether simply combining strong modules yields significant performance gains.
Despite these strong individual components, this combination significantly underperforms \texttt{HELIOS} by 19.3\% in AR@k and 35.4\% in nDCG@50. 
This highlights that the key novelty of \texttt{HELIOS} lies in how it addresses the inherent limitations of each retrieval module and integrates them using different levels of granularity—rather than merely stacking the methods together.
}


{\feditcolor
Table~\ref{tab:mmqa_retrieval_accuracy} shows the retrieval accuracy and end-to-end QA performance of \texttt{HELIOS} and \texttt{COS} on the \texttt{MMQA} development set.
\texttt{HELIOS} maintains its superior performance, achieving an average improvement of 20.9\% in AR across all $k$ values.
To further assess robustness, we measured end-to-end QA accuracy using GPT-4o as the reader with each method’s top-50 retrieved edges as input. 
\texttt{HELIOS} demonstrates substantial gains over \texttt{COS}, with improvements of 9.6\% in EM and 8.5\% in F1.
We claim that this result indicates the robustness of \texttt{HELIOS} across different datasets.
}

\begin{table}
    \centering
    \resizebox{0.75\columnwidth}{!}
    {%
        \begin{tabular}{l|cc|cc}
            \hline
            \multicolumn{1}{c|}{\multirow
            {2}{*}{\textbf{Algorithm}}} 
                                & \multicolumn{2}{c|}{\textbf{Dev}} & \multicolumn{2}{c}{\textbf{Test}} \\
            \cline{2-5}
                                & \textbf{EM}  & \textbf{F1} & \textbf{EM}  & \textbf{F1} \\
            \hline
            \texttt{OTTeR}               & 37.1              & 42.8                  & 37.3              & 43.1              \\
            \texttt{DoTTeR}              & 37.8              & 43.9                  & 35.9              & 42.0              \\
            \texttt{CORE}                & 49.0              & 55.7                  & 47.3              & 54.1              \\
            \texttt{COS}              & 56.9              & 63.2                  & 54.9              & 61.5              \\
            \hline
            \texttt{HELIOS}       & \textbf{59.3}     & \textbf{65.8}         & \textbf{57.0}     & \textbf{64.3}     \\
            \hline
        \end{tabular}
    }
    \vspace{-3mm}
    \caption{
        End-to-end QA accuracy on \texttt{OTT-QA}.
    }
    \label{tab:reading_accuracy} 
    \vspace{-3mm}
\end{table}

Table~\ref{tab:reading_accuracy} shows the end-to-end QA accuracy of \texttt{HELIOS} and \texttt{COS} on \texttt{OTT-QA}'s development and test sets.
Following \texttt{COS}, we used a \texttt{Fusion-in-Encoder (FiE)} model (\citealp{FiE}) fine-tuned on the \texttt{OTT-QA} dataset. 
For a fair comparison, we provided the reader with 50 edges, matching the number of edges used in \texttt{COS}.
The results indicate that compared to the \texttt{COS} model, our approach improved both EM and F1 scores by 4.2\% and 4.1\% on the development set, as well as by 3.8\% and 4.6\% on the test set, respectively.

\begin{figure}[t]
    \centering
    \small
    \begin{tabular}{c@{}c@{}}
        \includegraphics[width=0.5\columnwidth]{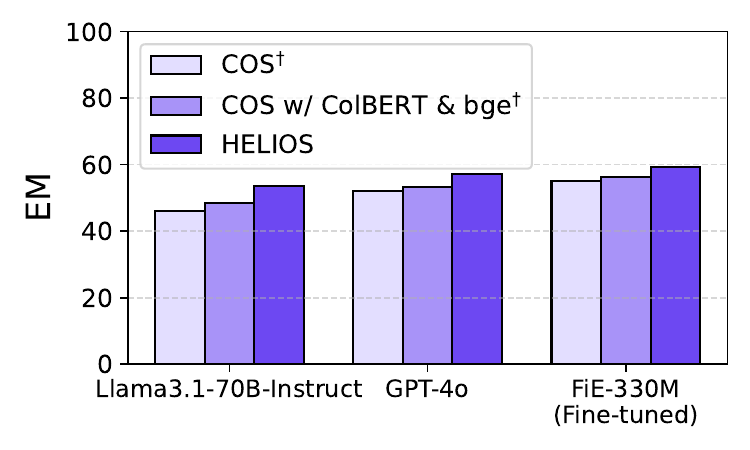} & 
        \includegraphics[width=0.5\columnwidth]{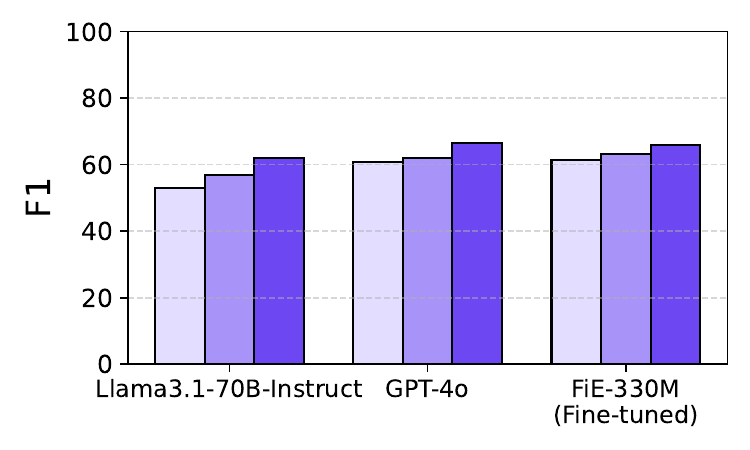} \\
        \noalign{\vskip -1mm}
        (a) EM Score & (b) F1 Score \\
    \end{tabular}
    \vspace{-3mm}
    \caption{
        End-to-end QA accuracy comparison across different readers for dev set of \texttt{OTT-QA}
    }
    \label{fig:reading_accuracy_across_diff_readers}
\end{figure}

\begin{table}[t]
    \centering
    \resizebox{0.5\columnwidth}{!}
    {%
        \begin{tabular}{l|cc}
            \hline
            \textbf{Algorithm} & \textbf{EM} & \textbf{F1} \\
            \hline
            \texttt{HOLMES}$^{\dagger}$    & 30.3        & 42.0        \\
            \texttt{HELIOS}    & \textbf{57.1} & \textbf{66.4} \\
            \hline
        \end{tabular}
    }
    \vspace{-2mm}
    \caption{
        QA Accuracy Comparison with \texttt{HOLMES} in a distractor setting adapted from \texttt{OTT-QA}.
    }
    \label{tab:holmes_comparison}
    \vspace{-3mm}
\end{table}

{\editcolor Figure~\ref{fig:reading_accuracy_across_diff_readers} presents the end-to-end QA accuracy of \texttt{HELIOS}, \texttt{COS}, and \texttt{COS w/ ColBERT \& bge} across different reader models, including \texttt{Llama-3.1-70B-Instruct} (\citealp{dubey2024llama}) and \texttt{GPT-4o} (\citealp{hurst2024gpt}).
For each algorithm, we report the results of the value of $k$ that yields the highest performance, where $k\in\{2, 5, 10, 20, 50\}$.
Detailed results for other values of $k$ are provided in the Appendix~\textsection~\ref{sec:reading_accuracy_supp}.
\texttt{HELIOS} improved the performance of all readers, achieving an average EM score improvement of 7.5\% and an average F1 score increase of 6.6\% compared to \texttt{COS w/ ColBERT \& bge}.
Through these results, we claim that our well-retrieved documents are capable of enhancing the effectiveness of various readers.
Examples of the prompts for the readers are in Appendix~\textsection~\ref{sec:Prompts}.}

{
\feditcolor
We also compare \texttt{HELIOS} with \texttt{HOLMES}, a graph-based multi-hop QA method tailored for distractor settings, to ensure a fair evaluation against existing graph-based approaches.
Since \texttt{OTT-QA} natively supports only the open-domain setting, we adapt it by constructing, for each question, a candidate set consisting of 10 items: the gold table, the gold passage, and the top eight hard negatives retrieved by \texttt{HELIOS}.

The results in Table~\ref{tab:holmes_comparison} show that \texttt{HELIOS} outperforms \texttt{HOLMES} by 88.4\% in EM and 58.1\% in F1.
We attribute this substantial performance gap to the following key factors:
(1) Seed Node Selection: \texttt{HOLMES} selects initial nodes based solely on named entities, which can overlook relevant, non-entity-centric context. 
\texttt{HELIOS}, by contrast, employs an edge-level, multi-vector retrieval strategy, capturing finer-grained relationships between tables and passages and leading to more accurate seed node retrieval.
(2) Question-Conditioned LLM Inference: \texttt{HOLMES} extracts triples via an LLM without conditioning on the question, often missing critical details. 
\texttt{HELIOS} leverages the reasoning ability of its LLM at inference time, ensuring that only query-relevant information is extracted and processed.
(3) Preserving Table Structure: \texttt{HOLMES} does not retain the structural nuances of tables when extracting triples, which hinders table-specific reasoning (e.g., column-wise aggregation). 
\texttt{HELIOS}, however, explicitly performs column-wise aggregation in its SLR module, supporting more complex tabular reasoning.
}

\vspace{-1mm}
\subsection{Ablation Study}
\label{sec:exp_ablation_study}
\vspace{-1mm}

\begin{table}[t]
    \centering
    \setlength{\tabcolsep}{2pt} 
\resizebox{\columnwidth}{!}{%
    \begin{tabular}{l|ccccc|c|cc}
        \hline
        \textbf{Algorithm}      & \textbf{AR@2}     & \textbf{AR@5}     & \textbf{AR@10}    & \textbf{AR@20}    & \textbf{AR@50}    & \textbf{nDCG@50}  & \textbf{EM} & \textbf{F1}   \\
        \hline
        \texttt{HELIOS}  & \textbf{63.3}     & 76.7     & 85.0     & \textbf{90.4}     & 94.2              & 47.0 & 59.3 & 65.8 \\
        \texttt{w/ Finetuned SLR}           & 63.2 & \textbf{76.8} & \textbf{85.1} & 90.3 & \textbf{94.8} & \textbf{47.6}  & \textbf{59.4} & \textbf{65.9}   \\
        \hline
        \texttt{w/o QNE}                 & 62.5              & 74.7              & 82.7              & 88.4              & 92.7              & 45.1        & 56.9 & 63.2      \\
        \texttt{w/o SLR}                 & 60.0              & 75.2              & 84.7              & 90.1              & 94.6     & 46.5   & 59.0  & 65.7            \\
        \hline
    \end{tabular}
    }
    \vspace{-3mm}
    \caption{\label{tab:ablation_study}
        Retrieval accuracy of \texttt{OTT-QA}'s dev set for \texttt{HELIOS}'s design factors.
    }
    \vspace{-3mm}
\end{table}

We performed an ablation study to assess the contribution of query-relevant node expansion (QNE) and star-based LLM refinement (SLR) to retrieval accuracy.
In \texttt{w/o QNE}, we removed the QNE and \texttt{HELIOS} passes the candidate bipartite subgraph $G_c$ directly to the SLR.
In \texttt{w/o SLR}, the SLR was removed and \texttt{HELIOS} decomposes the expanded graph $G_l$ into a list of edges.

As in Table~\ref{tab:ablation_study},
we found that removing the QNE module led to an average performance degradation of 2.1\% in AR across all $k$ values and 4.2\% in nDCG@50.
Additionally, excluding the QNE module resulted in a 4.2\% decrease in EM and a 4.1\% decrease in F1 scores.
This highlights QNE’s role in generating query-relevant edges missed by offline entity linking.
For the \texttt{w/o SLR}, we observed a noticeable drop in AR@2, AR@5, AR@10, AR@20, nDCG@50, EM score, and F1 score, with accuracy decreases of 5.5\%, 2.0\%, 0.4\%, 0.3\%, 1.1\%, 0.5\%, and 0.2\%, respectively.
This suggests that SLR helps accurately identify query-relevant nodes in complex queries requiring logical inference, particularly when $k$ is small.
Interestingly, for AR@50, \texttt{w/o SLR} slightly outperformed \texttt{HELIOS} by 0.4\%, likely due to LLM hallucinations.
Specifically, we observed 12 instances where the LLM failed to select the correct passage despite being provided with the ground truth passage.
We present the qualitative analysis results in Appendix \textsection~\ref{sec:qualitative_analysis}.

To mitigate hallucinations, we fine-tuned the SLR module using a training dataset synthesized with GPT-4o, incorporating tasks such as aggregate query detection, column-wise aggregation, and passage verification. 
In addition, we modified the prompt to favor false positives over false negatives by explicitly instructing the model to list passages even if only partially relevant. 
These strategies enabled the fine-tuned SLR to correctly identify the passage in 11 out of 12 previously failed instances. 
In the remaining case, the model selected the appropriate passage, but it was missing from the provided annotations. Further details are provided in Figure~\ref{fig:qualitative_analysis}(c) of Appendix~\textsection\ref{sec:qualitative_analysis}, and the fine-tuning process is described in Appendix~\textsection\ref{sec:star_based_llm_refinement_finetune}.


\vspace{-1mm}
\subsection{Impact of Granularity to Accuracy}
\label{sec:exp_granularity_analysis}
\vspace{-1mm}

\begin{table}[t]
    \centering
    \setlength{\tabcolsep}{2pt} 
    \resizebox{\columnwidth}{!}{%
        \begin{tabular}{l|ccccc|c}
        \hline
        \textbf{Retrieval Unit}  & \textbf{AR@2} & \textbf{AR@5}  & \textbf{AR@10} & \textbf{AR@20} & \textbf{AR@50} & \textbf{nDCG@50}\\
        \hline
        Node            & 29.3      & 47.4      & 58.8      & 68.5      & 79.5      & 23.8      \\
        Star Graph      & 37.9      & 57.4      & 66.9      & 76.4      & 84.5      & 28.5      \\
        Edge            & \textbf{49.1}      & \textbf{63.1}      & \textbf{70.6}      & \textbf{77.6}      & \textbf{85.1}      & \textbf{34.2}      \\
        \hline
    \end{tabular}
    }
    \vspace{-3mm}
    \caption{
        Retrieval accuracy across different units
    }
    \label{tab:retrieval_unit} 
\end{table}

We analyzed the impact of retrieval unit granularity on accuracy by comparing three versions of our subgraph retriever:
(i) Node retriever: Retrieves table segments first, then links related passages via entity linking.
(ii) Star graph retriever: Retrieves star graphs and integrates them into a graph.
(iii) Edge retriever: Retrieves edges and integrates them to construct a graph.
To ensure a fair comparison, we used the \texttt{ColBERTv2} baseline without fine-tuning.

As shown in Table~\ref{tab:retrieval_unit}, edge-based retrieval consistently outperformed the others. On average across all $k$ values, it exceeded star graph- and node-based retrieval by 6.9\% and 12.4\%, respectively, and for nDCG@50, by 20\% and 43.7\%. 
This highlights edge-based retrieval's ability to provide richer information while minimizing information loss, striking an effective balance compared to the others.

We further evaluated two refinement strategies using an LLM: a full graph prompt versus individual prompts for each star graph. The latter, which reduced irrelevant information in prompts, improved nDCG@50 by 12.4\% over the full graph setting (41.8), reducing hallucinations risks.

{\editcolor 
\vspace{-1mm}
\subsection{Algorithm Execution Time}
\label{sec:exp_efficiency}

\begin{table}[t]
    \centering
    \resizebox{\columnwidth}{!}{%
    \begin{tabular}{l|c|c}
        \hline
        \textbf{Algorithm}      & \textbf{Execution Time (s)}     & \textbf{nDCG@50}  \\
        \hline
        \texttt{DoTTeR} & 0.08    & 26.7 \\
        \texttt{CORE}  & 4.13     & 25.4 \\
        \texttt{COS}  & 3.75     & 33.6 \\
        \texttt{COS w/ ColBERT \& bge}  & 5.46     & 36.5 \\
        \texttt{HELIOS}  & 5.14     & 47.0 \\
        \texttt{HELIOS w/ Finetuned SLR} & 4.76 & \textbf{47.6} \\
        \hline
        \texttt{w/o SLR}                 & 2.16             & 46.5 \\
        \texttt{w/o (SLR \& Edge Reranker)} & \textbf{1.11} & 42.1 \\
        \hline
    \end{tabular}
    }
    \vspace{-3mm}
    \caption{\label{tab:computational_overhead}
        Algorithm execution time comparison
    }
\end{table}


As shown in Table~\ref{tab:computational_overhead}, \texttt{HELIOS} finds a sweet spot between the increase in algorithm runtime (1.37$\times$) and the increase in retrieval accuracy (39.9\% nDCG@50).
We further found out that fine-tuning LLM used in the SLR module can reduce the algorithm runtime to 1.26$\times$ that of \texttt{COS}, while boosting nDCG@50 to 41.7\%. 
Interestingly, runtime of \texttt{HELIOS} can be reduced to 0.57$\times$ that of \texttt{COS} by removing the SLR module, yet it shows a 38.4\% nDCG@50 gain.
We attribute this to \texttt{HELIOS} employing a beam search algorithm (beam width $b=10$), whereas \texttt{COS} performs expanded query retrieval on all 200 retrieved table segments, corresponding to have a beam width of 20 times larger size.
}

{\feditcolor 
To provide a clearer view of \texttt{HELIOS}’s accuracy–efficiency trade-off, we additionally compare it against \texttt{DoTTeR}, a single-vector retriever fine-tuned directly on \texttt{OTT-QA}.
While \texttt{DoTTeR} achieves end‑to‑end latency below 0.1s, its nDCG@50 score is markedly lower than that of \texttt{HELIOS}.
To push HELIOS’s end‑to‑end latency toward the one‑second mark, we ran an additional ablation in which we removed the SLR module and, in a second step, the edge re‑ranker. 
The streamlined variant processes a query in just 1.11s, a delay short enough that users perceive the system as “instant,” yet it still delivers a 57.7\% nDCG@50 gain over \texttt{DoTTeR}.
}

\vspace{-1mm}
\section{Conclusion}
\label{sec:conclusion}
\vspace{-1mm}

We presented \texttt{HELIOS}, a novel table-text retrieval method that harmonizes the strengths of both early and late fusion techniques while incorporating LLM reasoning. 
It addresses the limitations of competitors by introducing a multi-granular retrieval system that optimally balances granularity across retrieval stages. 
Experiments on OTT-QA show that it surpasses SOTA models, achieving a 42.6\% AR@2 improvement and a 39.9\% nDCG@50 gain.

{\editcolor 
\vspace{-1mm}
\section{Limitations}
\label{sec:limitations}
\vspace{-1mm}

Our approach currently focuses on the connections between table segments and passages. 
{\feditcolor
In future work, we aim to extend our method to more general types of connections between nodes of arbitrary modalities, such as images and inter-table links. 
Concretely, we are actively exploring extensions to HELIOS’s graph construction and traversal mechanisms to incorporate image-based nodes alongside text and tables. 
At this stage, we convert image inputs into text representations using vision-language models (e.g., Qwen2.5-VL 7B), and treat the resulting summaries as nodes. 
Even with this simple transformation, preliminary experimental results indicate that HELIOS already outperforms existing multimodal SOTA retrievers. We plan to further develop this line of research and present it in a dedicated future publication.
Moreover, as shown in Section~\ref{sec:exp_ablation_study}, the effectiveness of our SLR module can be limited by LLM hallucinations. 
An interesting direction for future work is to explore self-evaluation techniques~\citep{wangself, zhang2024self}—for instance, computing confidence scores and re-checking low-confidence outputs—to further reduce any remaining hallucinations.
}
}
{\editcolor 
\vspace{-1mm}
\section{Acknowledgments}
\label{sec:ack}
\vspace{-1mm}

This work was partly supported by Institute of Information \& communications Technology Planning \& Evaluation (IITP) grant funded by the Korea government (MSIT) (No. RS-2024-00509258, Global AI Frontier Lab, 70\%, No. RS-2024-00454666, Developing a Vector DB for Long-Term Memory Storage of Hyperscale AI Models, 15\%) and Basic Science Research Program through the National Research Foundation of Korea Ministry of Education(No. RS-2024-00415602, 15\%).
}

\bibliography{custom}

\appendix
\newpage
\section{Pseudocode and Schematic Workflow}
\setlength{\textfloatsep}{0pt}
\begin{algorithm}[h]
\small
\SetAlgoLined
\DontPrintSemicolon

\KwIn{Query $q$; early-fused graph $G_d$; edge retriever $f_e$; edge reranker $g_e$;
      node reranker $g_n$; expanded-query retrievers
      $f_{P\!\rightarrow S},\,f_{S\!\rightarrow P}$;
      star-based LLM Refiner $\mathcal{R}_\star$; beam width $b$; final list size $k$.}
\KwOut{Ranked edge list $E_q$.}

$\mathcal{E}_{\text{cand}} \leftarrow f_e(q,G_d)$ \tcp*[l]{retrieve top-$k_1$ edges}
$\mathcal{E}_{\text{cand}} \leftarrow g_e(q,\mathcal{E}_{\text{cand}})$ \tcp*[l]{rerank $\!\rightarrow$ top-$k_2$}
$G_c \leftarrow (V_c,\mathcal{E}_{\text{cand}})$ \tcp*[l]{candidate subgraph}

\BlankLine
$S \leftarrow \text{top-}b$ nodes by $g_n(q,V_c)$ \tcp*[l]{seed selection}
\ForEach{$u\in S$}{
  $q_u \leftarrow [\,q;\Gamma(u)\,]$\;
  \uIf{$u$ is a passage}{
    $N_u \leftarrow f_{P\!\rightarrow S}(q_u,G_d)$
  }\Else{
    $N_u \leftarrow f_{S\!\rightarrow P}(q_u,G_d)$
  }
  Add edges $\{(u,v)\mid v\in N_u\}$ to $G_c$\;
}

$G_\ell \leftarrow G_c$\;
isAgg $\leftarrow \mathcal{R}_\star.\textsc{DetectAgg}(q)$\;

Decompose $G_\ell$ into stars $\{\mathcal{S}_i\}$\;

$G_q \leftarrow \varnothing$\;
\ForEach{$\mathcal{S}_i$}{
  $G_q \leftarrow G_q \,\cup\,
          \mathcal{R}_\star.\textsc{Refine}\!\bigl(\mathcal{S}_i,\mathcal{G}_d,\,
          \texttt{agg}=\,\text{isAgg}\bigr)$\;
}
$E_q \leftarrow$ top-$k$ edges in $G_q$ (sorted by score)\;
\Return $E_q$
\caption{\textsc{HELIOS} Inference Pipeline}
\label{alg:helios_inference}
\end{algorithm}

\begin{figure*}[t]
\centering
    \includegraphics[width=\textwidth]{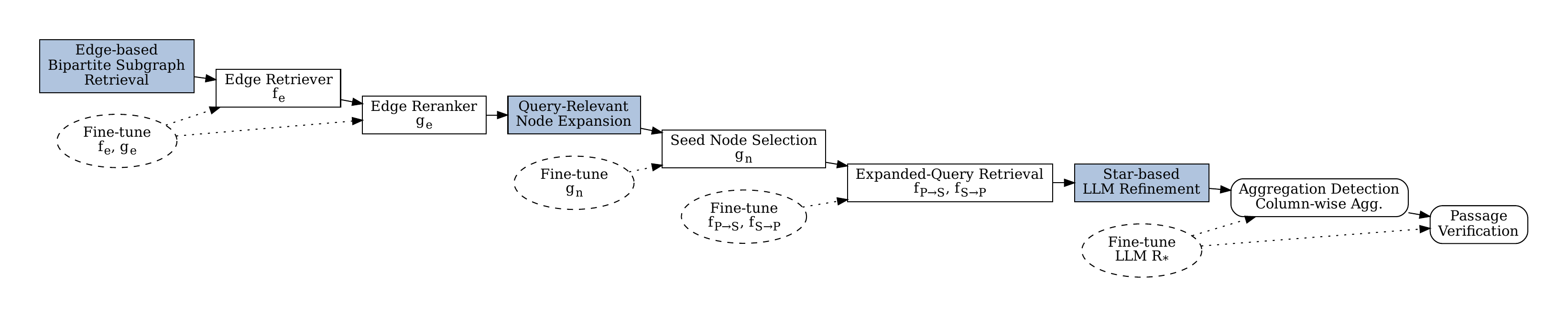}
    \caption{
        High-level schematic workflow of \texttt{HELIOS}.
    }
\label{fig:helios_workflow}
\end{figure*}
{
\feditcolor
To enhance the clarity of our framework, we provide both a pseudocode and a schematic workflow. 
The pseudocode in Algorithm~\ref{alg:helios_inference} details the step-by-step inference process, encompassing edge-based bipartite subgraph retrieval, query-relevant node expansion, and star-based LLM refinement. 
Additionally, Figure~\ref{fig:helios_workflow} offers a high-level schematic view of the overall workflow.
}







\section{Training Scheme}
\label{sec:training_scheme}

\subsection{Edge Retriever and Reranker}
\label{sec:edge_retriever_and_reranker}

The training scheme for our encoder $f_{e}$ follows the methodology outlined in \texttt{ColBERTv2} (\citealt{ColBERTv2}), leveraging a combination of in-batch negative loss and knowledge distillation loss to train the model. 
Specifically, the in-batch negative loss treats the edges corresponding to other queries within the same batch as negative samples.
This approach calculates a contrastive loss between the positive and negative edges.
In constructing the training dataset, it is crucial to have both positive and negative edges for each query. 
To define the positive edge, we use passages containing the answer and the associated table segments as ground truth and denoted as $x_{gt}$.
Conversely, negative edges are constructed by combining hard negative tables and passages from prior work (\citealt{COS}) with in-batch negative edges and are denoted as $n(q)$.
The contrastive loss $L_{cl}$ is represented as follows:
\begin{equation}
\small
    L_{cl} = - \sum_{(q,x_{gt})}\log{\frac{exp(s(q,x_{gt}))}{exp(s(q,x_{gt}))+\sum_{z\in n(q)}exp(s(q,z))}}
\end{equation}
\label{eqn:contrastive_loss}
The knowledge distillation process refines the edge encoder using a teacher-student model setup.
The distillation loss is computed based on the KL divergence between the score distribution generated by the teacher model and the training encoder. The training was conducted for 1 epoch with a batch size of 512 and a learning rate of 1e-5. We employed a cosine learning rate scheduler with 40 warm-up steps.

Here, the teacher model is the all-to-all interaction reranker $g_{e}$ fine-tuned with the contrastive loss, which serves as a more precise reference for edge relevance.
This method ensures that the encoder learns from a more sophisticated model, improving its capacity to accurately rank edges based on the query. The training was conducted for 1 epoch with a batch size of 96 and a learning rate of 2e-4. We employed a cosine learning rate scheduler with a warmup ratio of 0.1.

\subsection{Node Reranker}
\label{sec:node_reranker}

The training method for the node reranker $g_n$ is identical to that of the edge reranker $g_e$. 
For constructing the training dataset, we utilize the \texttt{OTT-QA} dataset \citep{OTT-QA}. 
Positive nodes are defined as those directly connected to the nodes that contain the correct answer in \texttt{OTT-QA}. 
In contrast, negative nodes are selected from the set of nodes retrieved through edge-based bipartite subgraph retrieval, excluding any nodes connected to the answer-containing nodes. The training was conducted for 1 epoch with a batch size of 96 and a learning rate of 2e-4. We used a cosine learning rate scheduler with a warm-up ratio of 0.1.

\subsection{Expanded Query Retrievers}
\label{sec:expanded_query_retrievers}

The training scheme for our expanded query retrievers $f_{S\rightarrow P}$, $f_{P\rightarrow S}$ also follow the methodology outlined in \texttt{ColBERTv2} (\citealt{ColBERTv2}).
To construct the training dataset, we generated triples consisting of the expanded query, positive node, and negative node. 
Expanded queries were created by incorporating nodes that are connected to the node containing the answer. 
Positive nodes consist of the nodes that contain the answer. Negative nodes are constructed using hard negative nodes as outlined in prior work (\citealt{COS}). The training was conducted for 1 epoch with a batch size of 512 and a learning rate of 1e-5. We employed a cosine learning rate scheduler with a 40 warmup steps.


\subsection{Star-based LLM Refinement}
\label{sec:star_based_llm_refinement_finetune}
We fine-tuned the Llama-3.1-Instruct 8B model \cite{dubey2024llama} using an instruction dataset synthesized with GPT-4o \cite{hurst2024gpt}. The dataset includes tasks for aggregate query detection, column-wise aggregation, and passage verification, and training was conducted for 1 epoch with a batch size of 128 and a learning rate of 2e-5. We employed a linear learning rate scheduler with a warmup ratio of 0.03.

To construct the training data, we applied edge-based bipartite graph retrieval and query-relevant node expansion on the OTT-QA training set, providing the results to GPT-4o for output generation. The final dataset was formed by combining these outputs with zero-shot prompts. Due to API cost constraints, we sampled 25\% of the OTT-QA training dataset, resulting in 1,655 samples each for query detection and column-wise aggregation and 7,010 samples for passage verification.

\section{Star-based LLM Refinement Supplementary}
\label{sec:star_based_llm_refinement_supp}
\begin{figure*}
\centering
    \includegraphics[width=\textwidth]{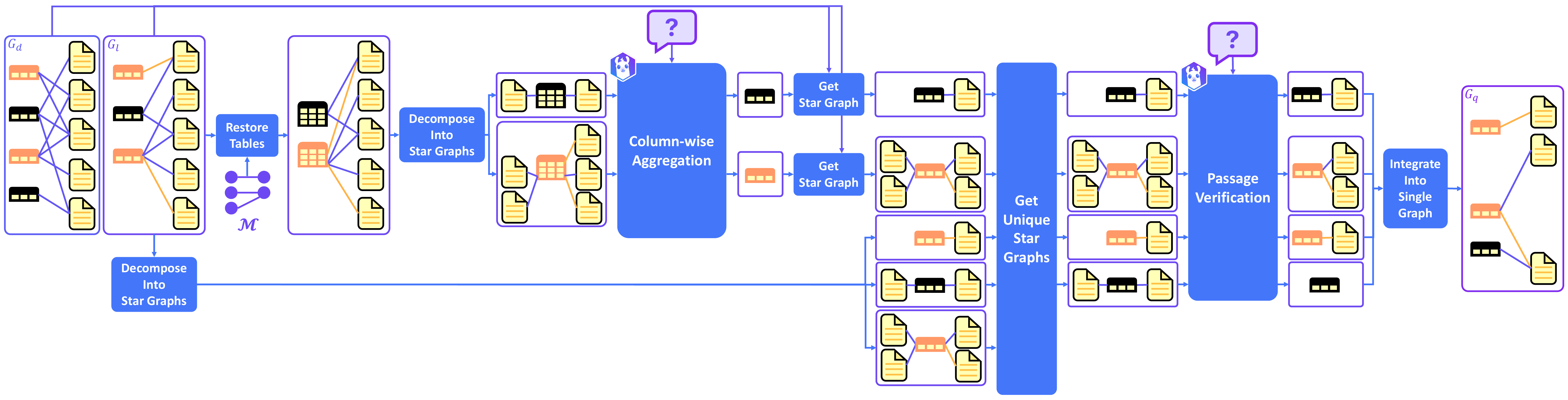}
    \caption{
        The overall process of star-based LLM refinement for queries classified as aggregation queries. 
        Table segment nodes of the same color (black, orange) indicate segments that belong to the same original table.
    }
\label{fig:star_based_llm_refinement}
\end{figure*}

In the Star-based LLM Refinement stage, the expanded graph from the Query-relevant Node Expansion (QNE) step undergoes further enhancement through the reasoning capabilities of LLMs. This stage consists of two primary operations: (1) aggregation of query-relevant table segments and (2) verification of query-irrelevant passages. The detailed process is illustrated in Figure~\ref{fig:star_based_llm_refinement}.

\section{Experiment Supplementaries}
\label{sec:experiment_supp}

\subsection{Implementation Details}
In our edge generation step (\textsection~\ref{sec:candidate_bipartite_graph_generation}), we used the same named entity recognition and entity linking models used by \texttt{COS} (\citealp{COS}). 
For the late-interaction edge retriever $f_e$ (\textsection~\ref{sec:candidate_bipartite_graph_generation}) and the expanded query retrievers $f_{P \rightarrow S}$ and $f_{S \rightarrow P}$ (\textsection~\ref{sec:candidate_bipartite_graph_augmentation}), we employed \texttt{ColBERTv2} (\citealp{ColBERTv2}) as the baseline model.
For the all-to-all interaction edge reranker $g_e$ (\textsection~\ref{sec:candidate_bipartite_graph_generation}) and node reranker $g_n$ (\textsection~\ref{sec:candidate_bipartite_graph_augmentation}), we used the \texttt{bge-reranker-v2-minicpm-layerwise} (\citealp{bge-reranker-v2-minicpm-layerwise}), specifically utilizing layer 24 as the baseline model. 
Lastly, for star-based LLM refinement (\textsection~\ref{sec:graph_refinement_via_complex_reasoning}), we used \texttt{Llama-3.1-8B-Instruct} (\citealp{dubey2024llama}) as the large language model.
In our experiments, the value of $k_1$ for the edge retriever $f_e$ was set to 400. 
Since \texttt{COS} selects the top-200 nodes as seed nodes, we fixed $k_2$ for the edge reranker $g_e$ to 100 to ensure a fair comparison.
We evaluated \texttt{HELIOS} on the OTT-QA's development set (2,214 examples) using four A100-80GB GPUs, which required approximately 3.4 hours.

\subsection{Parameter Sensitivity of Retriever}
\label{sec:exp_parameter_sensitivity}

\begin{figure}
\centering
    \includegraphics[width = \columnwidth]{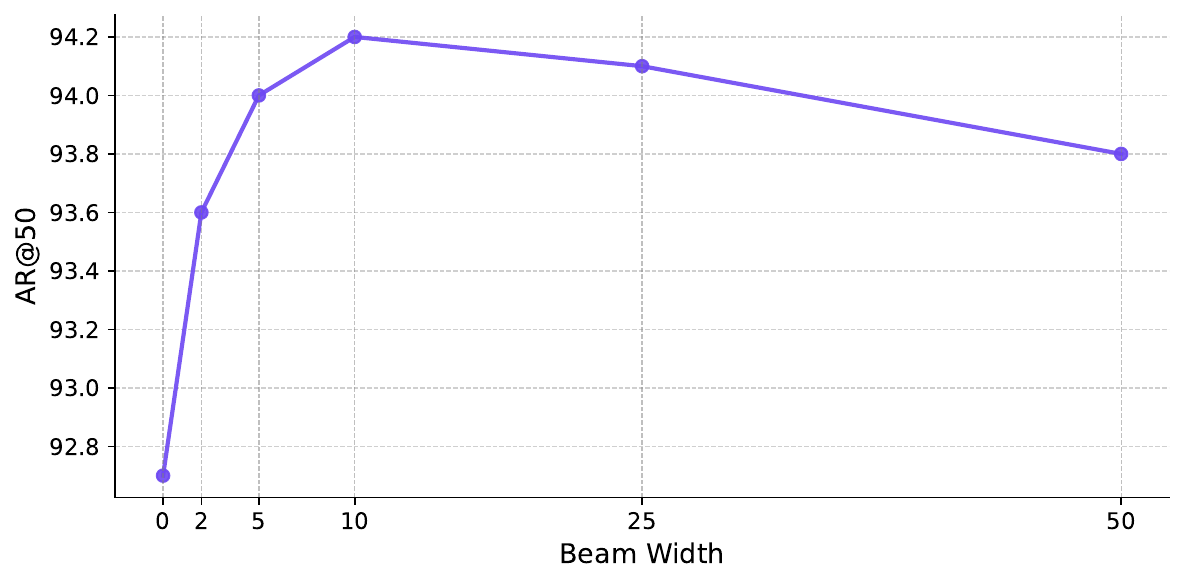}
    \caption{Change in AR@50 with varying beam width}
\label{fig:beam_width}
\end{figure}

We explored the impact of varying beam width $b$ on retrieval accuracy in terms of AR@50. 
The beam width directly influences the number of expanded nodes (\textsection~\ref{sec:candidate_bipartite_graph_augmentation}).
We experimented with beam widths of 0, 2, 5, 10, 25, 50 and measured the corresponding changes in AR@50.

Figure~\ref{fig:beam_width} illustrates the change in AR@50.
We observed that AR@50 was improved by 1.7\% as the beam width monotone increased from 0 to 10.
This indicates that larger beam widths lead to more accurate node augmentations by performing a more exhaustive search across the expanding node space. 
Interestingly, when the beam size increased to 50, AR@50 decreased slightly by 0.4\% compared to beam size 10. 
This drop may be due to LLM hallucinations in the star-based LLM refinement (SLR) module, where irrelevant edges were added to $G_l$, causing the SLR to fail in selecting the correct query-relevant nodes.
This observation highlights the importance of selectively expanding only the most probable nodes within the query-relevant node expansion module.

\subsection{Parameter Sensitivity of Reader}
\label{sec:reading_accuracy_supp}
\begin{figure}[t]
    \centering
    \small
    \begin{tabular}{c@{}c@{}}
        \includegraphics[width=0.5\columnwidth]{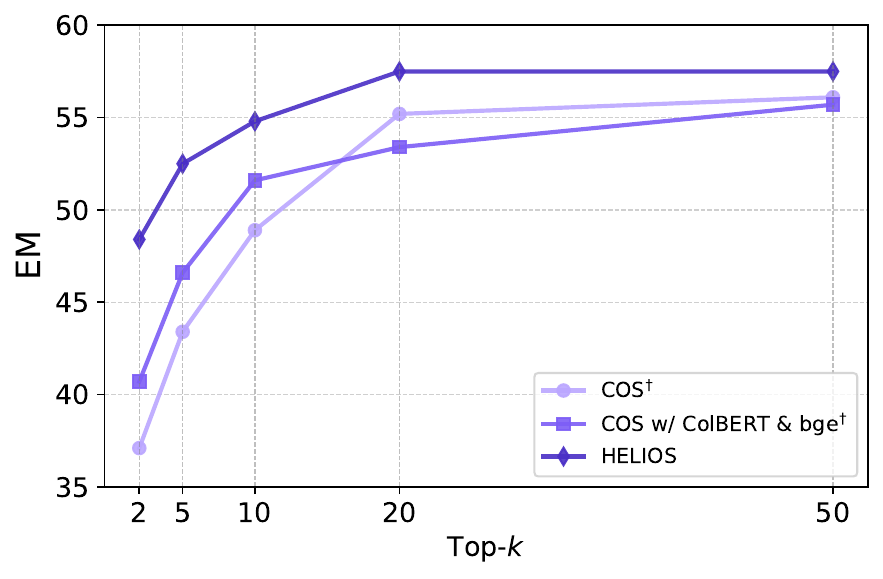} & 
        \includegraphics[width=0.5\columnwidth]{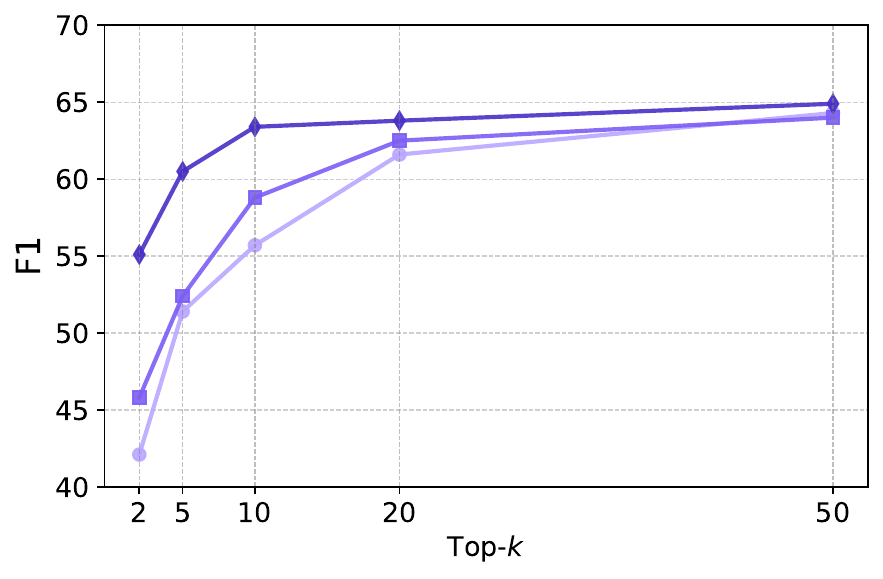} \\
        \noalign{\vskip -1mm}
        \multicolumn{2}{c}{(a) GPT-4o} \\
        \includegraphics[width=0.5\columnwidth]{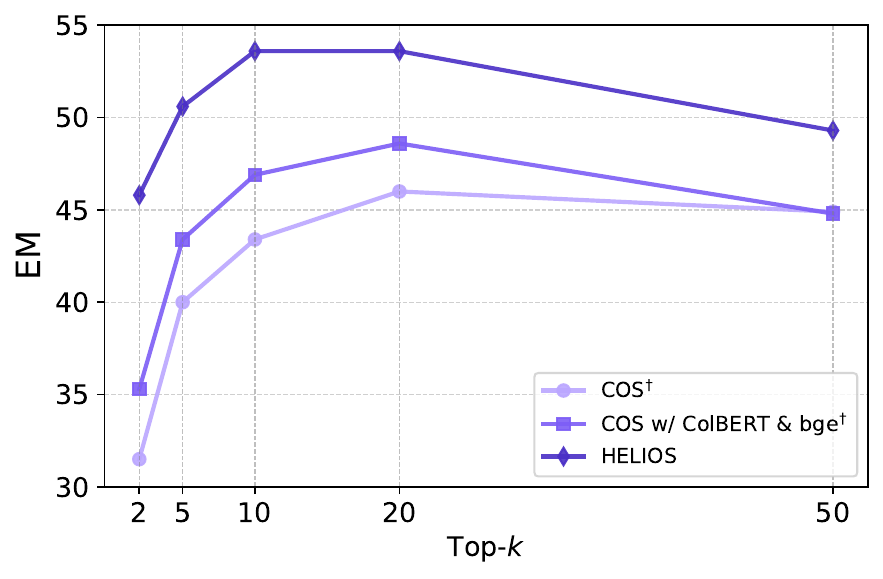} & 
        \includegraphics[width=0.5\columnwidth]{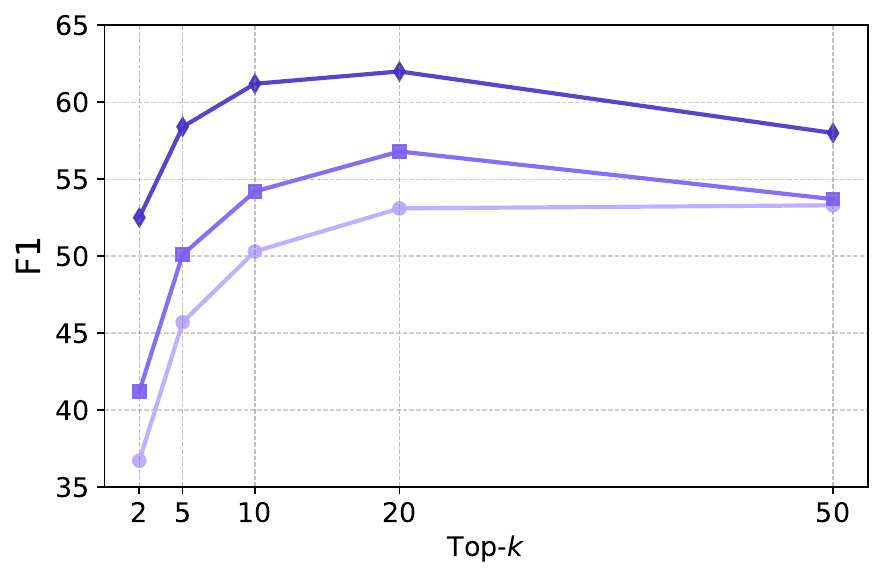} \\
        \noalign{\vskip -1mm}
        \multicolumn{2}{c}{(b) Llama3.1-Instruct-70B} \\
    \end{tabular}
    \vspace{-3mm}
    \caption{
        End-to-end QA accuracy comparison across different top-$k$ values on the OTT-QA dev set.
    }
    \label{fig:em_f1_comparison_over_diff_k}
\end{figure}
Figure~\ref{fig:em_f1_comparison_over_diff_k} presents the end-to-end QA accuracy of \texttt{HELIOS}, \texttt{COS}, and \texttt{COS w/ ColBERT \& bge} across different reader models—\texttt{Llama-3.1-70B-Instruct} (\citealp{dubey2024llama}) and \texttt{GPT-4o} (\citealp{hurst2024gpt})—with varying top-$k$ values ($k \in {2, 5, 10, 20, 50}$). 
Due to budget constraints, we sampled 10\% (221 out of 2,214) of the development set's QA pairs when evaluating accuracy variations for \texttt{GPT-4o}. 
As a result, \texttt{HELIOS} consistently achieves the highest AR@$k$ across all $k$ values. 
It improves the average EM score by 15.5\% for \texttt{Llama-3.1-70B-Instruct} and 9.2\% for \texttt{GPT-4o} compared to \texttt{COS w/ ColBERT \& bge}.

\subsection{Error and Reader Impact Analysis}
\label{sec:reader_impact}
{
\feditcolor
While \texttt{HELIOS} achieves high retrieval accuracy, there remains room for improvement in end-to-end QA performance. 
To investigate the gap between retrieval and QA accuracy, we performed an error analysis on 408 OTT-QA questions for which \texttt{HELIOS} successfully retrieved the gold evidence but still produced incorrect answers (F1 = 0).
We then re-tested these questions using \texttt{DeepSeek R1}~\citep{liu2024deepseek}, a stronger long-context reasoning model, with the same retrieved evidence.
Remarkably, the average F1 improved from 0 to 29.8, confirming that a more capable reader can better leverage the available evidence, especially for multi-hop questions, and further reduce hallucinations.
These findings highlight the potential for additional gains if a more advanced QA model is used on top of \texttt{HELIOS}’s strong retrieval.
}

\subsection{Detailed Breakdown of Offline Overhead}
\label{sec:appendix_offline_overhead}

{\feditcolor
Our method involves an offline stage for graph construction and retriever fine-tuning. 
The entire offline pipeline took approximately 49 hours, with graph construction accounting for about 38 hours. 
We did not optimize this phase since it is not the focus of our work. 
However, this stage can be significantly accelerated via straightforward parallelization (e.g., multi-core or multi-node setups), meaning the overhead does not compromise the feasibility of our approach.

Table~\ref{tab:graph_construction_time} provides a detailed breakdown of the graph construction process, which alone accounted for about 38 hours. 
In this phase, we construct a cross-modal retrieval graph by linking 400K tables and 5M passages via entity recognition and linking, following the approach of COS~\citep{COS}.

\begin{table}[h]
    \centering
    \setlength{\tabcolsep}{2pt}
    \resizebox{0.8\columnwidth}{!}{%
        \begin{tabular}{l|r}
        \hline
        \textbf{Component} & \textbf{Processing Time} \\
        \hline
        Entity recognition & 3h 14m 32s \\
        Entity linking & 2h 20m 48s \\
        Edge indexing & 22h 29m 54s \\
        Passage indexing & 7h 49m 01s \\
        Table segment indexing & 2h 29m 09s \\
        \hline
        \end{tabular}
    }
    \vspace{-3mm}
    \caption{Graph construction time by component.}
    \label{tab:graph_construction_time}
\end{table}

Table~\ref{tab:retriever_finetuning_time} summarizes the time required to fine-tune each retriever and reranker submodule used in our system. 
This phase took around 10 hours in total and can be parallelized with multi-core or multi-node setups. 
Given that both graph construction and retriever fine-tuning are offline processes, the overhead does not hinder the practicality of our system.

\begin{table}[h]
    \centering
    \setlength{\tabcolsep}{2pt}
    \resizebox{\columnwidth}{!}{%
        \begin{tabular}{l|r}
        \hline
        \textbf{Component} & \textbf{Training Time} \\
        \hline
        Edge retriever & 2h 49m 37s \\
        Edge reranker & 1h 25m 32s \\
        Node reranker & 1h 39m 20s \\
        Expanded query retriever (passages) & 36m 32s \\
        Expanded query retriever (tables) & 4h 14m 45s \\
        \hline
        \end{tabular}
    }
    \vspace{-3mm}
    \caption{Fine-tuning time for retriever submodules.}
    \label{tab:retriever_finetuning_time}
\end{table}
}

\begin{figure*}
\centering
    \includegraphics[width = \textwidth]{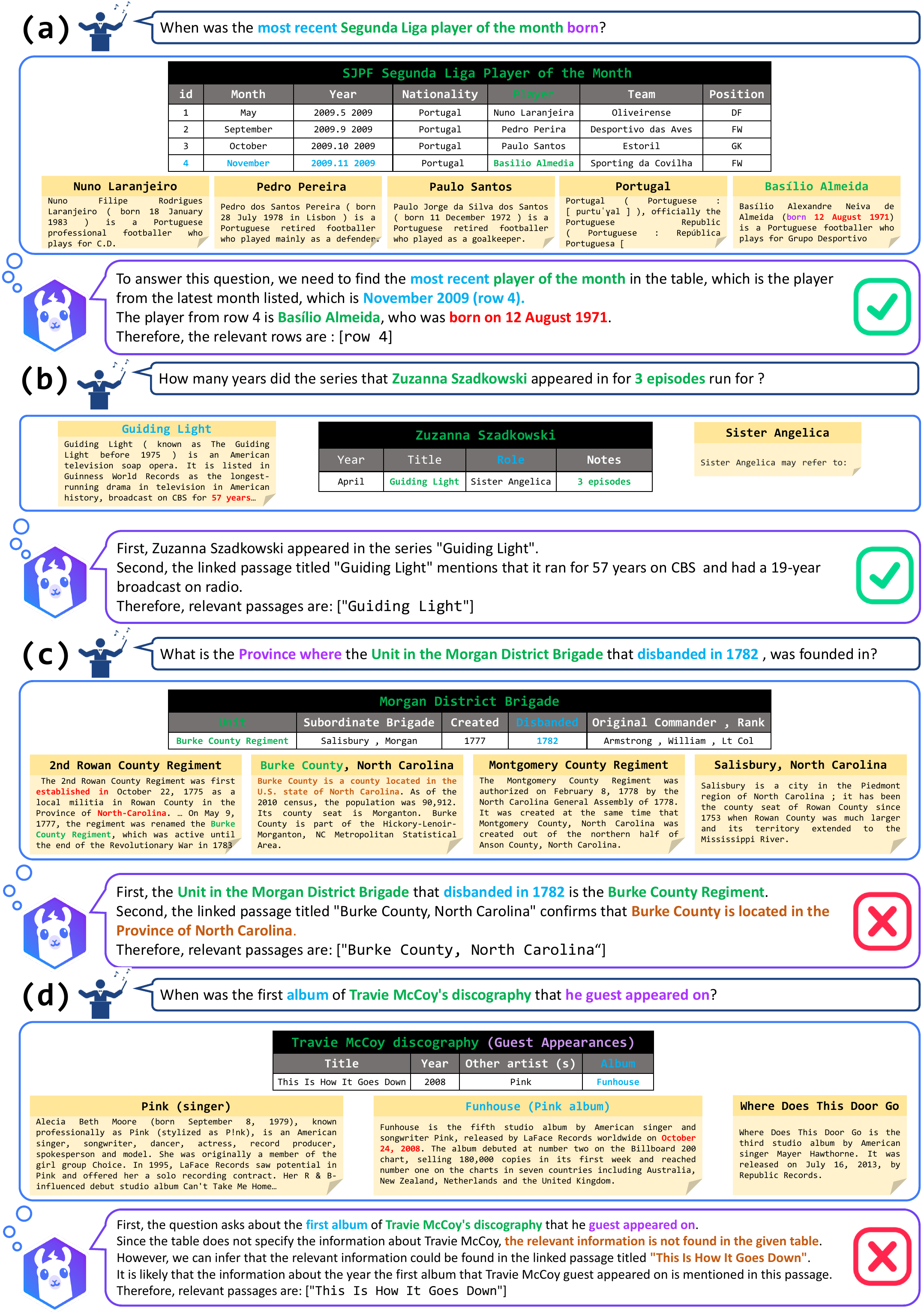}
    \caption{
        Qualitative analysis on four question-answer pairs.
        (a) A case where passage verification is successful.
        (b) A first case where passage verification has failed.
        (c) A second case where passage verification has failed.
        (d) A case where table aggregation is successful.
    }
\label{fig:qualitative_analysis}
\end{figure*}

\section{Qualitative Analysis}
\label{sec:qualitative_analysis}

In this section, we present a qualitative analysis of HELIOS's Column-wise Aggregation module and Passage Verification module, with the results illustrated in Figure~\ref{fig:qualitative_analysis}.
The subfigures in Figure~\ref{fig:qualitative_analysis} showcase the performance and distinctive scenarios for each module: 
(a) highlights successful cases of the Column-wise Aggregation module, while (b), (c), and (d) demonstrate representative cases related to the Passage Verification module. 
For each subfigure, the query is depicted in dark blue, the data provided to the sub-module are shown in light blue, and the inference result from the LLM is encapsulated in a purple speech bubble with a llama icon.

Figure~\ref{fig:qualitative_analysis}(a) shows a successful case of the column-wise aggregation module in resolving a complex query: identifying the birth date of the "most recent Segunda Liga Player of the Month." 
The essential part of answering this question was to recognize that the most recent player, Basilio Almeida, was honored in November 2009, as indicated in the SJPF Segunda Liga Player of the Month table. 
However, the initial data lacked the table segment containing the relevant row. 
The column-wise aggregation module reconstructed the table as shown in Figure~\ref{fig:qualitative_analysis}(a) to include this missing information, enabling the system to restore the row with the necessary details. 
The LLM correctly inferred from the reconstructed table that the row corresponding to the most recent player was Row 4, based on the Year and Month columns. 
This led \texttt{HELIOS} to accurately generate the final answer in this question, which is "12 August 1971."

Figure~\ref{fig:qualitative_analysis}(b) shows a successful case of the passage verification module in addressing the query, "How many years did the series that Zuzanna Szadkowski appeared in for 3 episodes run for?".
The module was provided with a Zuzanna Szadkowski table summarizing her appearances and a set of associated passages. 
The "Notes" column of the table segment confirmed that she appeared in three episodes of the series Guiding Light. 
The module correctly identified the one mentioning Guiding Light among the provided passages, the one which indicated that the series was broadcast on CBS for 57 years. 
the module correctly verified that the passage using the passage's information noting its broadcast duration, leading to an accurate answer.

Figure~\ref{fig:qualitative_analysis}(c) shows a failure case of the Passage Verification module when answering the query, "What is the province where the unit in the Morgan District Brigade that disbanded in 1782 was founded?".
The module correctly identified 'Burke County Regiment' as relevant to the query by recognizing from the 'Morgan District Brigade' table segment that the 'Disbanded' column value was 1782. 
However, information related to this query was present in two passages: '2nd Rowan County Regiment' and 'Burke County, North Carolina'. 
The LLM incorrectly verified only 'Burke County, North Carolina' as relevant, likely due to its more plausible-sounding title, while overlooking the correct answer 'North-Carolina' in the passage titled '2nd Rowan County Regiment'.
Consequently, the system produced an incorrect response, 'North Carolina'.
This error highlights two problems: (i) a limitation of the LLM reasoning capability and (ii) an example case of the \texttt{OTT-QA} benchmark's wrong answer annotation.

Figure 6(d) shows another failure case of the passage verification module, this time for the query, "When was the first album of Travie McCoy's discography that he guest appeared on?".
Prior retrieval results correctly introduced the ground truth table titled 'Travie McCoy discography (Guest Appearances)' to the passage verification module.
However, the LLM incorrectly inferred that "the table does not specify the information about Travie McCoy" as seen in the second line of its response bubble.
It then relied on its parameterized knowledge to wrongly verify a passage titled 'This Is How It Goes Down as relevant'. 
The correct answer 'Funhouse (Pink album)' was excluded from the final retrieved document set due to the verification error.

\section{Software and Data Licenses}
The licenses for the software and datasets used in this paper are as follows:
\begin{itemize}
    \item ColBERTv2: MIT License
    \item bge-reranker-v2-minicpm-layerwise: Apache 2.0 License
    \item LLaMA 3.1-8B-Instruct: LLaMA 3.1
    \item LLaMA 3.1-70B-Instruct: LLaMA 3.1
    \item Chain-of-Table: Apache 2.0 License
    \item TableLlama: MIT License
    \item COS: GPL-3.0 License
    \item OTT-QA: MIT License
\end{itemize}
All software and datasets were used strictly for research purposes and were not utilized in any non-research contexts, particularly for commercial applications.
The datasets used in this study, OTT-QA and MultimodalQA, consist of English-language data sourced from the Wikipedia domain.

\section{AI Assistants}
We used ChatGPT-4o (\citealp{hurst2024gpt}) to debug code efficiently, quickly identifying and resolving errors in our implementations. 
Additionally, we used it for rephrasing sentences in my writing to improve clarity and readability.

\section{Reproducibility Statement} 
OTTeR (\citealp{OTTeR}) and DoTTeR (\citealp{DoTTeR}) were reproduced using the official code available at \href{https://github.com/Jun-jie-Huang/OTTeR}{\texttt{OTTeR}} and \href{https://github.com/deokhk/DoTTeR}{\texttt{DoTTeR}}, respectively. COS (\citealp{COS}) and CORE (\citealp{CORE}) were reproduced using the official code from \href{https://github.com/Mayer123/UDT-QA}{\texttt{UDT-QA}}. 
The source code, data, and other artifacts for \texttt{HELIOS} have been made available at \href{https://github.com/pshlego/HELIOS}{\texttt{HELIOS}}.

\section{Prompt Templates}
\label{sec:Prompts}

For Star-based LLM refinement, we extended the prompt from \texttt{Chain-of-Table} (\citealp{chainoftable}), originally designed for selecting relevant rows from tables, to support column-wise aggregation and passage verification. 
This extension enables the joint consideration of table segments and linked passages. 
For the LLM reader, we constructed the prompt based on the HybridQA prompt from \texttt{TableLlama} (\citealp{tablellama}).

\newpage
\newpage


\begin{figure*}[htb]

\begin{tcolorbox}
[title = Aggregation Query Classification, colback = gray!10, colframe = black, sharp corners, boxrule=0.5mm]

Using \texttt{f\_agg()} API, return True to detect when a natural language query involves performing aggregation operations (max, min, avg, group by). 
Strictly follow the format of the below examples. Please provide your explanation first, then answer the question in a short phrase starting by ’Therefore, the answer is:’ \\
 \\
\textbf{Question}: when was the third highest paid Rangers F.C. player born? \\
\textbf{Explanation}: The question involves finding the birth date of the third highest paid player, which requires aggregation to find the third highest paid player. Therefore, the answer is: \texttt{f\_agg([True])} \\

\textbf{Question}: what is the full name of the Jesus College alumni who graduated in 1960? \\
\textbf{Explanation}: The question involves finding the full name of the alumni who graduated in 1960, which does not require aggregation. Therefore, the answer is: \texttt{f\_agg([False])} \\

\textbf{Question}: how tall, in feet, is the Basketball personality that was chosen as MVP most recently? \\
\textbf{Explanation}: The question involves finding the most recent MVP winner, which requires aggregation to identify the relevant player. Therefore, the answer is: \texttt{f\_agg([True])} \\

\textbf{Question}: what is the highest best score series 7 of Ballando con le Stelle for the best dancer born 3 July 1969? \\
\textbf{Explanation}: The question involves finding the highest score in a series for a specific dancer, which requires aggregation. Therefore, the answer is: \texttt{f\_agg([True])} \\

\textbf{Question}: which conquerors established the historical site in England that attracted 2,389,548 2009 tourists? \\
\textbf{Explanation}: The question involves identifying the conquerors who established a historical site, which does not require aggregation. Therefore, the answer is: \texttt{f\_agg([False])} \\

\textbf{Question}: what is the NYPD Blue character of the actor who was born on January 29, 1962? \\
\textbf{Explanation}: The question involves finding the character played by an actor born on a specific date, which does not require aggregation. Therefore, the answer is: \texttt{f\_agg([False])} \\
 \\
\textbf{Question}: `{\texttt{\{question\}}}' \\
\textbf{Explanation}: 

\end{tcolorbox}
\label{fig:agg_detection}
\end{figure*}

\newpage
\newpage


\begin{figure*}[htb]

\begin{tcolorbox}
[title = Column-wise Aggregation, colback = gray!10, colframe = black, sharp corners, boxrule=0.5mm]

Using \texttt{f\_row()} API to select relevant rows in the given table and linked passages that support or oppose the question. 
Strictly follow the format of the below example. Please provide your explanation first, then select relevant rows in a short phrase starting by: \textit{``Therefore, the relevant rows are:"}
\\
\\
\texttt{/* 
table caption : list of rangers f.c. records and statistics \\
col : \# | player | to | fee | date \\
row 1 : 1 | alan hutton | tottenham hotspur | 9,000,000 | 30 january 2008 \\
row 2 : 2 | giovanni van bronckhorst | arsenal | 8,500,000 | 20 june 2001 \\
row 3 : 3 | jean-alain boumsong | newcastle united | 8,000,000 | 1 january 2005 \\
row 4 : 4 | carlos cuellar | aston villa | 7,800,000 | 12 august 2008 \\
row 5 : 5 | barry ferguson | blackburn rovers | 7,500,000 | 29 august 2003 
*/}
 \\
\texttt{/* 
Passages linked to row 1: \\
- Alan Hutton: Alan Hutton (born 30 November 1984) is a Scottish former professional footballer, who played as a right back. Hutton started his career with Rangers, and won the league title in 2005. \\
- Tottenham Hotspur F.C.: Tottenham Hotspur Football Club, commonly referred to as Tottenham or Spurs, is an English professional football club in Tottenham, London, that competes in the Premier League.
*/}
 \\
\texttt{/* 
Passages linked to row 2: \\
- Giovanni van Bronckhorst: Giovanni Christiaan van Bronckhorst (born 5 February 1975), also known by his nickname Gio, is a retired Dutch footballer and currently the manager of Guangzhou R\&F. 
*/}
 \\
\texttt{/* 
Passages linked to row 3: \\
- Jean-Alain Boumsong: Jean-Alain Boumsong Somkong (born 14 December 1979) is a former professional football defender, including French international.  \\
- Newcastle United F.C.: Newcastle United Football Club is an English professional football club based in Newcastle upon Tyne, Tyne and Wear, that plays in the Premier League, the top tier of English football.
*/}
 \\
\textbf{Question:} '\texttt{When was the third highest paid Rangers F.C . player born ?}' \\
\textbf{Explanation:} The third-highest paid Rangers F.C. player, Jean-Alain Boumsong (row 3). \textit{Therefore, the relevant rows are:} \texttt{f\_row([row 3])}'
 \\
 \\
\texttt{/* '{\texttt{\{table\}}}' */} \\
 \\
\texttt{/* '{\texttt{\{linked\_passages\}}}' */} \\
 \\
\textbf{Question:} '{\texttt{\{question\}}}' \\
\textbf{Explanation:}

\end{tcolorbox}
\label{fig:column_aggregation}
\end{figure*}

\newpage
\newpage


\begin{figure*}[htb]

\begin{tcolorbox}
[title = Passage Verification, colback = gray!10, colframe = black, sharp corners, boxrule=0.5mm]

Using \texttt{f\_passage()} API to return a list of passage titles that are relevant to the question, even if they are only partially related. 
Strictly follow the format of the below example. Please provide your explanation first, then return a list of passages in a short phrase starting by: \textit{``Therefore, relevant passages are:"}
 \\
  \\
\texttt{/* 
table caption : List of politicians, lawyers, and civil servants educated at Jesus College, Oxford \\
col : Name | M | G | Degree | Notes \\
row 1 : Lalith Athulathmudali | 1955 | 1960 | BA Jurisprudence (2nd, 1958), BCL (2nd, 1960) | President of the Oxford Union (1958); a Sri Lankan politician; killed by the Tamil Tigers in 1993 
*/}
 \\
\texttt{/* 
List of linked passages: ["Law degree", "Oxford Union", "Lalith Athulathmudali"] \\
Title: Lalith Athulathmudali. Content: Lalith William Samarasekera Athulathmudali, PC (Sinhala; 26 November 1936 - 23 April 1993), known as Lalith Athulathmudali, was a Sri Lankan statesman. He was a prominent member of the United National Party, who served as Minister of Trade and Shipping; Minister of National Security and Deputy Minister of Defence; Minister of Agriculture, Food and Cooperatives, and finally Minister of Education. \\
Title: Law degree. Content: A law degree is an academic degree conferred for studies in law. Such degrees are generally preparation for legal careers; but while their curricula may be reviewed by legal authority, they do not themselves confer a license. A legal license is granted (typically by examination) and exercised locally; while the law degree can have local, international, and world-wide aspects. \\
Title: Oxford Union. Content: The Oxford Union Society, commonly referred to simply as the Oxford Union, is a debating society in the city of Oxford, England, whose membership is drawn primarily from the University of Oxford. Founded in 1823, it is one of Britain’s oldest university unions and one of the world’s most prestigious private students' societies. The Oxford Union exists independently from the university and is separate from the Oxford University Student Union.
*/}
 \\
  \\
\textbf{Question:} What is the full name of the Jesus College alumni who graduated in 1960? \\
\textbf{Explanation:} First, Lalith Athulathmudali graduated in 1960. Second, the linked passage titled ``Lalith Athulathmudali" confirms his full name. \textit{Therefore, relevant passages are:} \texttt{f\_passage(["Lalith Athulathmudali"])}
 \\
 \\
\texttt{/* '{\texttt{\{table\}}}' */} \\
 \\
\texttt{/* `{\texttt{\{linked\_passages\}}}' */} \\
 \\
\textbf{Question:} `{\texttt{\{question\}}}' \\
\textbf{Explanation:}

\end{tcolorbox}

\end{figure*}

\newpage
\begin{figure*}[htb]

\begin{tcolorbox}
[title = Hybrid Question Answering, colback = gray!10, colframe = black, sharp corners, boxrule=0.5mm]
Below is an instruction that describes a task, paired with an input that provides further context. Write a response that appropriately completes the request.
\\
\#\#\# Instruction: \\
This is a hybrid question answering task. \\
The goal of this task is to answer the question given tables and passages. \\
Strictly follow the format of the below examples. Please provide your explanation first, then answer the question in a short phrase starting by: \textit{`Therefore, the answer is:`}
 \\
\#\#\# Examples: \\
\texttt{
Title : List of politicians, lawyers, and civil servants educated at Jesus College, Oxford \\
col : Name | M | G | Degree | Notes \\
row 1 : Lalith Athulathmudali | 1955 | 1960 | BA Jurisprudence ( 2nd , 1958 ) , BCL ( 2nd , 1960 ) | President of the Oxford Union ( 1958 ) ; a Sri Lankan politician ; killed by the Tamil Tigers in 1993 \\
row 2 : Neal Blewett ( HF ) | 1957 | 1959 | BA PPE ( 2nd ) | Member of the Australian House of Representatives ( 1977-1994 ) , Government Minister ( 1983-1994 ) , High Commissioner to the UK ( 1994-1998 ) \\
row 3 : Joseph Clearihue | 1911 | 1914 | BA Jurisprudence ( 2nd , 1913 ) , BCL ( 3rd , 1914 ) | Canadian Rhodes scholar ; later became a member of the Legislative Assembly of British Columbia and a county court judge ; also chairman of the council of Victoria College , British Columbia ( which became the University of Victoria under his leadership ) \\
}
 \\
\texttt{ 
Passages linked to row 1: \\
- [Lalith Athulathmudali](https://en.wikipedia.org/wiki/Lalith\_Athulathmudali) Lalith William Samarasekera Athulathmudali , PC (26 November 1936 - 23 April 1993) , known as Lalith Athulathmudali , was Sri Lankan statesman . He was a prominent member of the United National Party , who served as Minister of Trade and Shipping ; Minister National Security and Deputy Minister of Defence. \\
Passages linked to row 3: \\
- [Joseph Clearihue](https://en.wikipedia.org/wiki/Joseph\_Clearihue) Joseph Badenoch Clearihue ( 1887-1976 ) was a Canadian lawyer , judge , academic and politician . \\
}
\\
\textbf{Question:}What is the full name of the Jesus College alumni who graduated in 1960? \\
\textbf{Explanation:} Lalith Athulathmudali graduated in 1960, and his full name is Lalith William Samarasekera Athulathmudali. Therefore, the answer is: Lalith William Samarasekera Athulathmudali.
\\
\#\#\# Input: \\
Here are the table and passages to answer this question. Please provide your explanation first, then
answer the question in a short phrase starting by \textit{`Therefore, the answer is:`}
\\
 \\
\texttt{/* `{\texttt{\{table\_and\_linked\_passages\}}}' */} \\
 \\
\textbf{Question:} `{\texttt{\{question\}}}' \\
\textbf{Explanation:}

\end{tcolorbox}
\label{fig:passage_verification}
\end{figure*}

\end{document}